\documentclass[epj]{svjour}
\usepackage{graphics}

\usepackage{epsfig,amsmath,amssymb} 
\usepackage{graphics}
\usepackage{multirow}
\usepackage{subfig}
\begin{document}
\title{Magnetic order in spin-1 and spin-$\frac{3}{2}$ interpolating square-triangle Heisenberg antiferromagnets}
%\subtitle{Magnetic order in $s>\frac{1}{2}$ interpolating square-triangle magnetic models}
\titlerunning{Magnetic order in $s>\frac{1}{2}$ interpolating square-triangle magnetic models}
%\titlerunning
\author{First author\inst{1} \and Second author\inst{2}% etc
\author{P.~H.~Y.~Li
\and R.~F.~Bishop}
}                     % Do not remove

%\author{P.~H.~Y.~Li\inst{1}
%\and R.~F.~Bishop\inst{1}
%\and D.~J.~J.~Farnell\inst{2}
%\and C.~E.~Campbell\inst{3}
%}                     % Do not remove
%
\offprints{}          % Insert a name or remove this line
%
%\institute{Insert the first address here \and the second here}
\institute{School of Physics and Astronomy, The University of Manchester, Schuster Building, Manchester, M13 9PL, United Kingdom}
%
%University of Glamorgan, Pontypridd CF37 1DL, Wales, United Kingdom
%\and School of Physics and Astronomy, University of Minnesota, 116 Church Street SE, Minneapolis, Minnesota 55455, USA}
%
\date{Received: date / Revised version: date}

\abstract{
%Insert your abstract here.
  Using the coupled cluster method we investigate spin-$s$ 
  $J_{1}$-$J_{2}'$ Heisenberg antiferromagnets
  (HAFs) on an infinite, anisotropic, two-dimensional triangular
  lattice for the two cases where the spin quantum number $s=1$ and
  $s=\frac{3}{2}$.  With respect to an underlying square-lattice geometry the
  model has antiferromagnetic ($J_{1} > 0$) bonds between nearest
  neighbours and competing ($J_{2}' > 0$) bonds between
  next-nearest neighbours across only one of the diagonals of each
  square plaquette, the same diagonal in each square.  In a
  topologically equivalent triangular-lattice geometry, the model has
  two types of nearest-neighbour bonds: namely the $J_{2}' \equiv
  \kappa J_{1}$ bonds along parallel chains and the $J_{1}$ bonds
  producing an interchain coupling.  The model thus interpolates
  between an isotropic HAF on the square lattice at one limit ($\kappa
  = 0$) and a set of decoupled chains at the other limit ($\kappa
  \rightarrow \infty$), with the isotropic HAF on the triangular
  lattice in between at $\kappa = 1$.  For both the spin-1 model and the 
  spin-$\frac{3}{2}$ model we find a second-order type of quantum phase
  transition at $\kappa_{c}=0.615 \pm 0.010$ and
  $\kappa_{c}=0.575 \pm 0.005$ respectively, between a N\'{e}el 
  antiferromagnetic state and a helically ordered state.  
  In both cases the ground-state energy $E$ and its 
  first derivative $dE/d\kappa$ are continuous at 
  $\kappa=\kappa_{c}$, while the order parameter for the
  transition (viz., the average ground-state on-site magnetization)
  does not go to zero there on either side of the transition.  The
  phase transition at $\kappa = \kappa_{c}$ between the N\'{e}el
  antiferromagnetic phase and the helical phase for both the $s=1$ and
  $s=\frac{3}{2}$ cases is analogous to that also observed in our previous 
  work for the $s=\frac{1}{2}$ case at a value $\kappa_{c}=0.80 \pm 0.01$.  However, 
  for the higher spin values the transition appears to be of continuous 
  (second-order) type, exactly as in the classical case, whereas for the 
  $s=\frac{1}{2}$ case it appears to be weakly first-order in nature (although 
  a second-order transition could not be ruled out entirely).
\PACS{
  %    {PACS-key}{discribing text of that key}   \and
 % {PACS-key}{discribing text of that key}
     {75.10.Jm}{Quantized spin models}   \and
      {75.30.Kz}{Magnetic phase boundaries} \and
      {75.50.Ee}{Antiferromagnetics} 
     } % end of PACS codes
} %end of abstract
\maketitle
%
%\section{Introduction}
%\label{intro}
%Your text comes here. Separate text sections with
%\section{Section title}
%\label{sec:1}
%and \cite{RefJ}
%\subsection{Subsection title}
%\label{sec:2}
%as required. Don't forget to give each section
%and subsection a unique label (see Sect.~\ref{sec:1}).
%\begin{figure}
%\resizebox{0.75\textwidth}{!}{%
%\vspace{5cm}       % Give the correct figure height in cm
%\end{figure}
%\begin{figure*}
%\vspace*{5cm}       % Give the correct figure height in cm
%\end{figure*}
%\begin{table}
%\begin{tabular}{lll}
%\hline\noalign{\smallskip}
%\end{tabular}
% Or use
%\vspace*{5cm}  % with the correct table height
%\end{table}

\section{Introduction}
\label{Intro}

In recent years, the theoretical study of two-dimensional (2D) quantum
spin systems has been intensely motivated by the fact that such models
often describe well the properties of real magnetic materials of great experimental
interest.  It is an encouraging fact that experiments have often
supported theoretical predictions or vice versa.
Moreover, the study of frustration and quantum fluctuation in quantum spin-lattice
systems has developed into an extremely active area of research.  The interplay
between frustration and quantum fluctuations in magnetic systems can 
produce a wide range of fascinating quantum phases
\cite{Sa:1995,Ri:2004,Mi:2005}.  Both effects are in principle capable of 
destabilising or completely destroying the magnetic order of the spin system.
In turn this might then lead to the formation of a spin-liquid phase or be 
responsible for other quantum phenomena of similar significant interest.
The driving force for the differentiated manifestation of the various 
kinds of possible quantum effects can, in principle, also come from the type and 
nature of the underlying crystallographic lattice, from the
number and variety of the magnetic bonds, and from the magnitude of the spin quantum
numbers of the atoms that reside on the atomic lattice sites
\cite{ccm_UJack_asUJ_2010}.  It is thus of considerable interest to try to study 
the effects of each of these driving forces in turn for 
specific model Hamiltonians.

A particularly well studied 2D model is the frustrated spin-$\frac{1}{2}$
$J_{1}$-$J_{2}$ model on the square lattice with nearest-neighbour
(NN) bonds ($J_{1}$) and next-nearest-neighbour (NNN) bonds ($J_{2}$),
for which it is now well accepted that in the case where both sorts of 
bonds are antiferromagnetic ($J_i >0; i=1,2$) there exist two antiferromagnetic phases
exhibiting magnetic long-range order (LRO) at small and
at large values of $\alpha \equiv J_{2}/J_{1}$ respectively.  These are separated
by an intermediate quantum paramagnetic phase without magnetic LRO in
the parameter regime $\alpha_{c_{1}} < \alpha < \alpha_{c_{2}}$, where
$\alpha_{c_{1}} \approx 0.4$ and $\alpha_{c_{2}} \approx 0.6$.  For
$\alpha < \alpha_{c_{1}}$ the ground-state (gs) phase exhibits N\'{e}el magnetic LRO,
whereas for $\alpha > \alpha_{c_{2}}$ it exhibits collinear stripe-ordered LRO.

As already noted above, the spin quantum number $s$ can, both in 
principle and in practice, play an important role in
the phase structure of strongly correlated spin-lattice systems, which often
display rich and interesting phase scenarios due to the interplay
between the quantum fluctuations and the competing interactions.
Varying the spin quantum number $s$ can tune the strength of the
quantum fluctuations and lead to fascinating phenomena 
\cite{Da:2005_JPhy_17}.  A well-known example of such spin-dependent behaviour
is the gapped Haldane phase \cite{Ha:1983} in $s=1$ one-dimensional
(1D) chains, which is not present in their $s=\frac{1}{2}$
counterparts.  

Some recent studies on large-spin (i.e., $s>\frac{1}{2}$) systems include:  (a) the
comparison of the Heisenberg antiferromagnet (HAF) on the Sierpi\'{n}ski 
gasket with the corresponding HAFs on various regular 2D lattices, including 
the square, honeycomb, triangular and kagom\'{e} lattices, 
for the cases $s=\frac{1}{2}$, 1, and $\frac{3}{2}$ \cite{Vo:2001}; (b) the one-dimensional (1D)
Heisenberg chain for both integral and half-odd integral values of $s$ 
up to a value of 10 \cite{Bi:1992,Fa:2002,Grover:2010_s1}; (c) the 2D 
$J$-$J'$ model on the square lattice containing two different types of 
NN bonds, for values of $s$ between  $\frac{1}{2}$ and $2$ \cite{Da:2005_JPhy_17}; 
d) the 2D spin-anisotropic $XXZ$ Heisenberg Hamiltonian for $s=1$ 
\cite{Bi:1992,Lin:1989,Ir:1992,Fa:2001_PRB64}; 
e) the spatially anisotropic $J_{1}$-$J_{1}$'-$J_{2}$ model 
on the 2D square lattice for values of $s$ 
up to $4$ \cite{Mo:2006,Bi:2008_EPL}; 
(f) the spin-anisotropic $J_{1}^{XXZ}$-$J_{2}^{XXZ}$ model for 
$s=1$ \cite{Bi:2008_JPCM_V20_p415213};
(g) the pure $J_{1}$-$J_{2}$ model for $s=1$ \cite{Ji:2009};
(h) the 2D Union Jack model for $s=1$ and
$s=\frac{3}{2}$ \cite{Bishop:2010_UJack_GrtSpins};  
and the 2D Heisenberg model on the honeycomb lattice for $s=1$ \cite{Zhao:2011_honeycomb_s1}.  
Another example that has been experimentally 
studied involves the investigation of the single-ion anisotropy energy in the 2D 
kagom\'{e} lattice for the case $s=\frac{5}{2}$ \cite{Vr:2008}.  

Also noteworthy in this context is the recent discovery of 
superconductivity with a transition
temperature at $T_{c} \approx 26\,$K in the layered iron-based compound
LaOFeAs, when doped by partial substitution of the oxygen atoms by
fluorine atoms~\cite{KWHH:2008}.  This finding has been followed by the rapid
discovery of superconductivity at even higher values of $T_{c}$
($\gtrsim 50\,$K) in a broad class of similar quaternary
compounds.  Enormous interest has thereby been engendered in this class
of materials.  The very recent first-principles
calculations~\cite{Ma:2008} shows, for example, that the undoped parent precursor
material LaOFeAs of the first material investigated in this oxypnictide class 
is well described by the spin-1 $J_{1}$-$J_{2}$
model on the square
lattice.

We have previously used the coupled cluster method
(CCM)~\cite{Bi:1991,Bi:1998,Fa:2004} to study the magnetic order in a
spin-half interpolating square-triangle HAF (viz., the
$J_{1}$-$J_{2}'$ model) \cite{Bi:2008_SqTrian,Bi:2010_SqTrian_IJMPB}.  
This is a variant of the well-known $J_{1}$-$J_{2}$ 
model on the infinite 2D square lattice, described below, in which one half of the NNN
$J_2$ bonds are removed.  In
the present paper we further the investigation of the $J_{1}$-$J_{2}'$
model by replacing the spin-$\frac{1}{2}$ particles by particles with
$s = 1$ and $s = \frac{3}{2}$.  The 2D spin-$\frac{1}{2}$
$J_{1}$-$J_{2}'$ model has also been studied recently by other
means~\cite{Me:1999,We:1999,St:2007,Pa:2008}, but we know of no other studies 
of the model for spins with spin quantum number $s>\frac{1}{2}$.

\section{The model}
\label{model_section}
\begin{figure}[t]
\includegraphics[width=8cm]{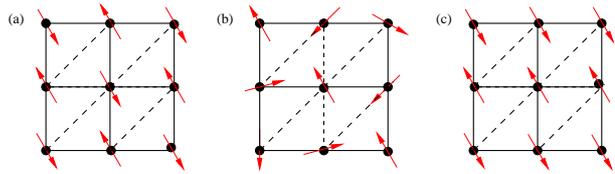}
\caption{(colour online) $J_{1}$-$J_{2}'$ model; -- $J_{1}$; - - -
  $J_{2}'$; (a) N\'{e}el state, (b) spiral state, (c) striped state.}
\label{model}
%\end{center}
\end{figure}

The Hamiltonian of the $J_{1}$-$J_{2}'$ model is written as
\begin{equation}
H = J_{1}\sum_{\langle i,j \rangle}{\bf s}_{i}\cdot{\bf s}_{j} + J_{2}'\sum_{[i,k]}{\bf s}_{i}\cdot {\bf s}_{k}  \label{H}
\end{equation}
where the operators ${\bf s}_{i} \equiv (s^{x}_{i}, s^{y}_{i},
s^{z}_{i})$ are the spin operators on lattice site $i$, with 
${\bf  s}^{2}_{i} = s(s+1)$, and we consider here the cases 
$s=1$ and $s=\frac{3}{2}$.  On the square lattice the sum
over $\langle i,j \rangle$ runs over all distinct NN bonds, but the
sum over $[i,k]$ runs only over one half of the distinct NNN bonds
with equivalent bonds chosen in each square plaquette, as shown
explicitly in Fig.~\ref{model}.
We shall be interested here only
in the case of competing (or frustrating) antiferromagnetic bonds
$J_{1} > 0$ and $J_{2}' > 0$, and henceforth for all of the results
shown we set $J_{1} \equiv 1$. Clearly, the model may be described
equivalently as a Heisenberg model on an anisotropic triangular
lattice in which each triangular plaquette contains two NN $J_{1}$
bonds and one NN $J_{2}'$ bond. The model thus interpolates
continuously between HAFs on a square lattice ($J_{2}' = 0$) and on a
triangular lattice ($J_{2}'=J_{1}$). Similarly, when $J_{1}=0$ (or
$J_{2}' \rightarrow \infty$ in our normalization with $J_{1} \equiv
1$) the model reduces to uncoupled 1D chains (along the chosen
diagonals on the square lattice). The case $J_{2}' \gg 1$ thus
corresponds to weakly coupled 1D chains, and hence the model 
also interpolates between 1D and 2D scenarios.  
As well as the obvious
theoretical richness of the model, there is also experimental interest
since it is also believed to well describe such quasi-2D crystalline materials as 
organic compounds containing BEDT-TTF ~\cite{Ki:1996}, for which with $J_{2}'/J_{1}$ 
lies typically in the range from about 0.3 to about 1; and
Cs$_{2}$CuCl$_{4}$~\cite{Co:1997}, for which $J_{2}'/J_{1}$ takes a value 
of about 6, thus making this material even quasi-1D.

The $J_{1}$-$J_{2}'$ model has only two gs phases in the classical 
case (corresponding to the limit where the spin quantum number $s
\rightarrow \infty$). For $J_{2}' < \frac{1}{2}J_{1}$ the gs phase is N\'{e}el
ordered, as shown in Fig.~\ref{model}(a), whereas for $J_{2}' >
\frac{1}{2}J_{1}$ it has spiral order, as shown in Fig.~\ref{model}(b),
wherein the spin direction at lattice site ($i,j$) points at an angle
$\alpha_{ij}=\alpha_{0}+(i+j)\alpha_{{\rm cl}}$, with $\alpha_{{\rm
    cl}}={\rm cos}^{-1}(-\frac{J_{1}}{2J_{2}'}) \equiv \pi -
\phi_{{\rm cl}}$. The pitch angle $\phi_{{\rm cl}}={\rm
  cos}^{-1}(\frac{J_{1}}{2J_{2}'})$ thus measures the deviation from
N\'{e}el order, and it varies from zero for $2J_{2}'/J_{1} \leq 1$ to
$\frac{1}{2} \pi$ as $J_{2}'/J_{1} \rightarrow \infty$, as shown later in
Fig.~\ref{angleVSj2}. When $J_{2}'=J_{1}$ we regain the classical
3-sublattice ordering on the triangular lattice with $\alpha_{{\rm
    cl}} = \frac{2}{3}\pi$. The classical phase transition at
$J_{2}'=\frac{1}{2}J_{1}$ is of continuous (second-order) type, with
the gs energy and its derivative both continuous.

In the limit of large $J_{2}'/J_{1}$ the above classical limit
represents a set of decoupled 1D HAF chains (along the diagonals of
the square lattice) with a relative spin orientation between
neighboring chains that approaches 90$^{\circ}$. In fact, of course,
there is complete degeneracy at the classical level in this limit
between all states for which the relative ordering directions of spins
on different HAF chains are arbitrary. Clearly the exact spin-$\frac{1}{2}$
limit should also be a set of decoupled HAF chains as given by the
exact Bethe ansatz solution~\cite{Be:1931}. However, one might expect
that this degeneracy could be lifted by quantum fluctuations by the
well-known phenomenon of {\it order by disorder}~\cite{Vi:1977}. Just
such a phase is known to exist in the $J_{1}$-$J_{2}$
model~\cite{Bi:2008_JPCM_V20_p255251,Bi:2008_PRB} for values of $J_{2}/J_{1}
\gtrsim 0.6$, where it is the so-called collinear striped phase in
which, on the square lattice, spins along (say) the rows in
Fig.~\ref{model} order ferromagnetically while spins along the columns
and diagonals order antiferromagnetically, as shown in Fig.\
\ref{model}(c). We investigate the possibility below whether a
stripe-ordered phase may be stabilized by quantum fluctuations at
larger values of $\kappa$ for either of the cases $s=1$ or
$s=\frac{3}{2}$, in order to compare with the earlier $s=\frac{1}{2}$
case for which we found \cite{Bi:2008_SqTrian} that such a gs phase 
might exist for high enough values of the frustration parameter $\kappa$, 
as discussed below.

Thus, for the $s=\frac{1}{2}$ case our own CCM
calculations~\cite{Bi:2008_SqTrian} provided strong evidence that
the spiral phase becomes unstable at large values of the
frustration parameter $\kappa$. In view of that observation we also
used the CCM for the $s=\frac{1}{2}$ case with the collinear
stripe-ordered state as a model state. We found
tentative evidence, based on the relative energies of the two states,
for a second zero-temperature phase transition between the spiral and
stripe-ordered states at a larger critical value of
$\kappa_{c_{2}} \approx 1.8 \pm 0.4$, as well as firm evidence for a
first phase transition between the N\'{e}el antiferromagnetic phase
and the helical phase at a critical coupling
$\kappa_{c_{1}} = 0.80 \pm 0.01$.

The transition at $\kappa
= \kappa_{c_{1}}$ for the $s=\frac{1}{2}$ case was found to be an
interesting one. As in the classical ($s \rightarrow \infty$) case,
the energy and its first derivative appeared to be continuous
(within the errors inherent in our approximations), thus providing a
typical scenario of a second-order phase transition, although a weakly
first-order one could not be excluded since the gs energy did in fact 
show some definite signs of a (weak) discontinuity in slope. Furthermore, the average
on-site magnetization was seen to approach a value $M_{c_{1}}
= 0.025 \pm 0.025$ very close to zero on both sides of the transition, 
but with a very sharp drop and hence a possible discontinuity in $M$ on the
spiral side of the transition, as is often more
typical of a first-order transition. 

A particular interest
here is to compare and contrast the corresponding transition(s)
between the $s=\frac{1}{2}$ and the $s>\frac{1}{2}$ models.
By contrast with the $s=\frac{1}{2}$ we shall find below that 
for the cases with $s=1$ and $s=\frac{3}{2}$ the average on-site magnetization
at the analogous phase transition between N\'{e}el-ordered and 
spirally-ordered states approaches smoothly the same nonzero value on both sides
of the transition.
Such continuous phase transitions where the order parameter
does not vanish are
well known in quantum magnetism. A prototypical example is the $s=\frac{1}{2}$ 
anisotropic $XXZ$ model with a Hamiltonian given by
%%%%%%%%%%%%%%%%
\begin{equation}
H=J \sum_{\langle i,j \rangle}(s^{x}_{i}s^{x}_{j} + s^{y}_{i}s^{y}_{j} + \Delta s^{z}_{i}s^{z}_{j}),  \label{H_XXZ}
\end{equation}
%%%%%%%%%%%%%%%
and which thus contains only NN anisotropic, antiferromagnetic ($J >
0$) Heisenberg bonds. Classically (corresponding to the $s \rightarrow
\infty$ limit) the model has a continuous phase transition at $\Delta
= \Delta_{c} \equiv 1$ between two different N\'{e}el
antiferromagnetic phases, one aligned along the $z$-axis for $\Delta >
1$, and the other along some arbitrary direction in the perpendicular
$xy$-plane for $-1 < \Delta < 1$. The $s=\frac{1}{2}$ model may also
be solved exactly on a 1D chain by the Bethe ansatz
technique~\cite{Orbach:1958_BetheAns}.  It is found in this 1D case
that as the critical point is approached, $\Delta \rightarrow
\Delta_{c} = 1$, from either side, the average on-site (or staggered)
magnetization $M \rightarrow M_{c} = 0$. The approach to zero is of a
quite nontrivial kind, via a function with an essential singularity at
$\Delta = \Delta_{c}$. By contrast, for the same $s=\frac{1}{2}$ $XXZ$
model of Eq.~(\ref{H_XXZ}) on a 2D square lattice, $M \rightarrow
M_{c} \approx 0.31$ as $\Delta \rightarrow \Delta_{c} = 1$. Thus, the
type of phase transition we observe below for the spin-$1$ and 
$s=\frac{3}{2}$ 2D interpolating square-triangle Heisenberg
antiferromagnets is quite analogous to the one at $\Delta = \Delta_{c}
= 1$ in the $s=\frac{1}{2}$ $XXZ$ model on the 2D square lattice, but
not to that of the same $XXZ$ model on the 1D chain.

We now first briefly describe the main elements of the CCM below in
Sec.~\ref{CCM}, where we also discuss the approximation schemes used
in practice for the $s=\frac{1}{2}$ case and the $s>\frac{1}{2}$
cases. Then in Sec.~\ref{Results} we present our CCM results based on
using the N\'{e}el, spiral and striped states discussed above as
model states (or starting states). We conclude in
Sec.~\ref{discussion} with a discussion of the results.

\section{The coupled cluster method}
\label{CCM}

The CCM (see, e.g., Refs.~\cite{Bi:1991,Bi:1998,Fa:2004} and
references cited therein) is regarded as one of the most powerful and
most versatile modern techniques in quantum many-body theory. It has
been successfully applied to many quantum magnets 
(see Refs.~\cite{ccm_UJack_asUJ_2010,Fa:2004,Bi:2008_JPCM_V20_p255251,Bi:2008_PRB,Ze:1998,Kr:2000,Fa:2001,Da:2005,Schm:2006,Bishop:2010_KagomeSq})
and references cited therein).  The CCM is suitable for studying
frustrated systems, for which the main alternative methods are often
only of limited usefulness. For example, quantum Monte Carlo
techniques are usually restricted by the sign problem for such
systems, and the exact diagonalization method is limited in practice,
especially for $s>\frac{1}{2}$, to such small lattices that it is
often insensitive to the details of any subtle phase order present.

The CCM method to solve the gs Schr\"{o}dinger ket and
bra equations, $H|\Psi\rangle = E|\Psi\rangle$ and
$\langle\tilde{\Psi}|H=E\langle\tilde{\Psi}|$ respectively is now 
briefly outlined (and see
Refs.~\cite{Fa:2002,Bi:1991,Bi:1998,Fa:2004,Ze:1998,Kr:2000}
for further details). The implementation of
the CCM is initiated by the selection of a model state $|\Phi\rangle$ on top of
which to incorporate later in a systematic fashion the multispin
correlations contained in the exact ground states $|\Psi\rangle$ and
$\langle\tilde{\Psi}|$. The CCM employs the exponential ansatz,
$|\Psi\rangle=e^{S}|\Phi\rangle$ and
$\langle\tilde{\Psi}|=\langle\Phi|\tilde{S}$e$^{-S}$. The creation 
correlation operator $S$ is written as 
$S = \sum_{I\neq0}{\cal S}_{I}C^{+}_{I}$ 
with its destruction counterpart as $\tilde{S} = 1 +
\sum_{I\neq0}\tilde{\cal S}_{I}C^{-}_{I}$. The operators $C^{+}_{I}
\equiv (C^{-}_{I})^{\dagger}$, with $C^{+}_{0} \equiv 1$, have the
property that $\langle\Phi|C^{+}_{I} = 0 = C^{-}_{0}|\Phi\rangle \,;\, \forall I \neq 0$. They
form a complete set of multispin creation operators with respect to
the model state $|\Phi\rangle$. The calculation of the ket- and
bra-state correlation coefficients 
$({\cal S}_{I}, \tilde{{\cal S}_{I}})$ 
is performed by requiring the gs energy expectation value
$\bar{H} \equiv \langle\tilde{\Psi}|H|\Psi\rangle$ to be a minimum
with respect to each of them. This results in a coupled set of
equations $\langle
\Phi|C^{-}_{I}\mbox{e}^{-S}H\mbox{e}^{S}|\Phi\rangle = 0$ and
$\langle\Phi|\tilde{S}(\mbox{e}^{-S}H\mbox{e}^{S} -
E)C^{+}_{I}|\Phi\rangle = 0\,;\, \forall I \neq 0$, which we normally
solve by using parallel computing routines \cite{ccm} for the
correlation coefficients $({\cal S}_{I}, \tilde{{\cal S}_{I}})$ within
specific truncation schemes as outlined below.
 
In order to treat each lattice site on an equal footing a mathematical
rotation of the local spin axes on each lattice site is performed such
that every spin of the model state aligns along its negative $z$-axis.
As a result, our description of the spins is given wholly in terms of
these locally defined spin coordinate frames. The multispin creation
operators may be expressed as \(C^{+}_{I}\equiv s^{+}_{i_{1}}
s^{+}_{i_{2}} \cdots s^{+}_{i_{n}}\), in terms of the locally defined
spin-raising operators $s^{+}_{i} \equiv s^{x}_{i} + s^{y}_{i}$ on
lattice sites $i$. Upon solving for the multispin cluster correlation
coefficients $({\cal S}_{I}, \tilde{{\cal S}_{I}})$ as outlined above,
the gs energy $E$ may then be calculated from the relation
$E=\langle\Phi|\mbox{e}^{-S}H\mbox{e}^{S}|\Phi\rangle$, and the gs
staggered magnetization $M$ from the relation $M \equiv -\frac{1}{N}
\langle\tilde{\Psi}|\sum_{i=1}^{N}s^{z}_{i}|\Psi\rangle$ in terms of
the rotated spin coordinates.

If a complete set of multispin configurations $\{I\}$ with respect to
the model state $|\Phi\rangle$ is included in the calculation of the
correlation operators $S$ and $\tilde{S}$, then the CCM formalism
becomes exact. However, it is necessary in practical applications to use systematic
approximation schemes to truncate them to some finite subset.  
For the $s=\frac{1}{2}$ case, the localised LSUB$n$ scheme is
commonly employed, as in our earlier
paper on the $s=\frac{1}{2}$ version of the present
model \cite{Bi:2008_SqTrian}, as well as in our other previous
work \cite{Fa:2002,Fa:2004,Ze:1998,Kr:2000,Schm:2006}. 
Under this truncation scheme 
all possible multi-spin-flip correlations over different locales on
the lattice defined by $n$ or fewer contiguous lattice sites are
retained.  A cluster is defined as having $n$ contiguous sites if 
every one of the $n$ sites is adjacent (as a nearest neighbour) to
at least one other.  Clearly this definition, however, depends  on 
how we choose the geometry of the lattice among various topologically 
equivalent possibilities that may exist.  For example, the current 
model may be construed as referring to sites on a square lattice, as
shown in Fig.~\ref{model}.  In this case the $J_{2}'$ bonds, for example, 
join NNN sites (which, by definition, are thus not adjacent).   
Alternatively, the model may be equivalently construed as referring to sites on a triangular 
lattice, in which case both $J_{1}$ and $J_{2}'$ bonds join NN (and 
hence adjacent) sites.  In all of the results presented here 
we consider the model to be defined on a triangular lattice in 
making CCM approximations.

However, we note that the number of fundamental LSUB$n$ configurations
for $s>\frac{1}{2}$ becomes appreciably higher than for
$s=\frac{1}{2}$, since each spin on each site $i$ can now be raised up to 
$2s$ times by the spin-raising operator $s^{+}_{i}$. Thus, for the
$s>\frac{1}{2}$ models it is more practical, but equally systematic, to
use the alternative SUB$n$-$m$ scheme, in which all correlations
involving up to $n$ spin flips spanning a range of no more than $m$
contiguous lattice sites are retained
\cite{Bi:2008_EPL,Bi:2008_JPCM_V20_p415213,Bishop:2010_UJack_GrtSpins,Fa:2004,Fa:2001}.
We then set $m=n$, and hence employ the so-called SUB$n$-$n$ scheme.
More generally, the LSUB$m$ scheme is thus equivalent to the
SUB$n$-$m$ scheme for $n=2sm$ for particles of spin $s$.  For
$s=\frac{1}{2}$, LSUB$n\equiv$ SUB$n$-$n$; whereas for
$s>\frac{1}{2}$, LSUB$n\equiv$ SUB2$sn$-$n$. The numbers of such
fundamental configurations (viz., those that are distinct under the
symmetries of the Hamiltonian and of the model state $|\Phi\rangle$)
that are retained for the N\'{e}el and striped model states of the current
$s=1$ and $=\frac{3}{2}$ models at various SUB$n$-$n$ levels, 
defined with respect to an underlying triangular-lattice geometry, are shown
in Table~\ref{FundConf_spin1_SUBnn}.
\begin{table}[!tbp]
\caption{Number of fundamental CCM configurations ($N_{f}$) for the SUB$n$-$n$ ($n=\{2,3,4,5,6,7\}$) 
scheme for the N\'{e}el and striped model states, for the $J_{1}$-$J_{2}'$ model defined on 
a triangular lattice, for the spin-1 and spin-$\frac{3}{2}$ cases.}
\label{FundConf_spin1_SUBnn}
\vskip0.5cm
\begin{tabular}{cccccc}
\hline\noalign{\smallskip}
%\multirow{2}{*}{Method} & \multicolumn{2}{c}{$s=1$} & &  \multicolumn{2}{c}{$s=\frac{3}{2}$} \\ %\cline{2-3} \cline{5-6}
%Method & \multicolumn{2}{c}{$s=1$} & &  \multicolumn{2}{c}{$s=\frac{3}{2}$} \\ %\cline{2-3} \cline{5-6}
\multirow{2}{*}{Method} & \multicolumn{2}{c}{$s=1$} & &  \multicolumn{2}{c}{$s=\frac{3}{2}$} \\ %\cline{2-3} \cline{5-6}
\noalign{\smallskip}\cline{2-3} \cline{5-6}\noalign{\smallskip}
%\noalign{\smallskip}\hline\noalign{\smaallskip}
 & \multicolumn{2}{c}{$N_{f}$} & & \multicolumn{2}{c}{$N_{f}$} \\ 
\noalign{\smallskip}\hline\noalign{\smallskip}
SUB$n$-$n$ & striped & spiral & & striped & spiral  \\ 
\noalign{\smallskip}\hline\noalign{\smallskip}
SUB$2$-$2$ & 2 & 4 & & 2 & 4 \\ 
SUB$3$-$3$ & 4 & 26 & & 4 & 27 \\ 
SUB$4$-$4$ & 60 & 189 & & 60 & 211 \\ 
SUB$5$-$5$ & 175 & 1578 & & 175 & 1908 \\ 
SUB$6$-$6$ & 2996 & 14084 & & 3622 & 18501 \\ 
SUB$7$-$7$ & 11778 & 131473 & & 13320 & 188326 \\ 
\noalign{\smallskip}\hline
\end{tabular} 
\end{table}

Although we never need to perform any finite-size scaling, since all
CCM approximations are automatically performed from the outset in the
$N \rightarrow \infty$ limit, where $N$ is the total number of 
lattice sites, we do need as a last step to extrapolate
to the $n \rightarrow \infty$ limit in the truncation index $n$. We
use here the
well-tested~\cite{Bi:2008_EPL,Bi:2008_JPCM_V20_p415213,Bi:2008_SqTrian,Kr:2000,Fa:2001}
empirical scaling laws
\begin{equation}
E/N=a_{0}+a_{1}n^{-2}+a_{2}n^{-4}\;,  \label{Extrapo_E}
\end{equation} 
\begin{equation}
M=b_{0}+b_{1}n^{-1}+b_{2}n^{-2}\;, \label{Extrapo_M}
\end{equation} 
exactly as we did previously for the corresponding $s=\frac{1}{2}$ model 
\cite{Bi:2008_SqTrian}, for the gs energy per
spin $E/N$ and the gs staggered magnetization $M$, respectively.

\section{Results}
\label{Results}
The results of the CCM calculations are reported here for the spin-1
and spin-$\frac{3}{2}$ $J_{1}$-$J_{2}'$ model Hamiltonian of Eq.\ (\ref{H}),
using the N\'{e}el, spiral and striped states shown in Fig, 1(a)-(c) 
as CCM model states, and with the SUB$n$-$n$ approximation scheme defined 
with respect to an underlying triangular-lattice geometry.  We
set the parameter $J_{1}=1$. Our available computational power 
at present is such that we can
perform SUB$n$-$n$ calculations for the spiral model state 
(viz., the state that requires the highest number of fundamental
configurations ($N_{f}$) for a given SUB$n$-$n$ truncation
index $n$) only for values $n \leq 7$ for both the $s=1$ 
and $s=\frac{3}{2}$ cases.  
We thus present results for each of the N\'{e}el, striped and spiral
states only up to the SUB$7$-$7$ level, for the sake of consistency 
in our extrapolations to the $n \rightarrow \infty$ limit . 

We note that, as
has been well documented in the past~\cite{Fa:2008}, the LSUB$n$ (or
SUB$n$-$n$) data for both the gs energy per spin $E/N$ and the average on-site
magnetization $M$ converge differently for even-$n$ sequences and
odd-$n$ sequences, similar to what is frequently observed in
perturbation theory~\cite{Mo:1953}. Since, as a general rule, it is
desirable to have at least ($n+1$) data points to fit to any fitting
formula that contains $n$ unknown parameters, we prefer to have at
least 4 results to fit to Eqs.\ (\ref{Extrapo_E}) and
(\ref{Extrapo_M}). Both the available odd and even series of our SUB$n$-$n$
data violate this desirable rule. However, our results (for both sets $n=\{2,4,6\}$ 
and $n=\{3,5,7\}$) for the $s=\frac{1}{2}$ case are consistent 
with those using the larger LSUB$m$ sequences available in this case.  
This gives us confidence in both the accuracy of our results and the robustness
of our extrapolation schemes. Hence, for most of our extrapolated
results below we use the even SUB$n$-$n$ sequence with $n=\{2,4,6\}$
and the odd SUB$n$-$n$ sequence with $n=\{3,5,7\}$.

Firstly, the results obtained using the spiral model state are
reported.  For this state we first perform CCM calculations with
the pitch angle $\phi$ as a free parameter.  At each separate level of approximation 
we then choose the angle $\phi = \phi_{{\rm SUB}n-n}$ that minimizes the energy 
$E_{{\rm SUB}n-n}(\phi)$.  

\begin{figure*}[!p]
\begin{center}
\mbox{
\subfloat[$s=1$]{\scalebox{0.3}{\includegraphics[angle=270]{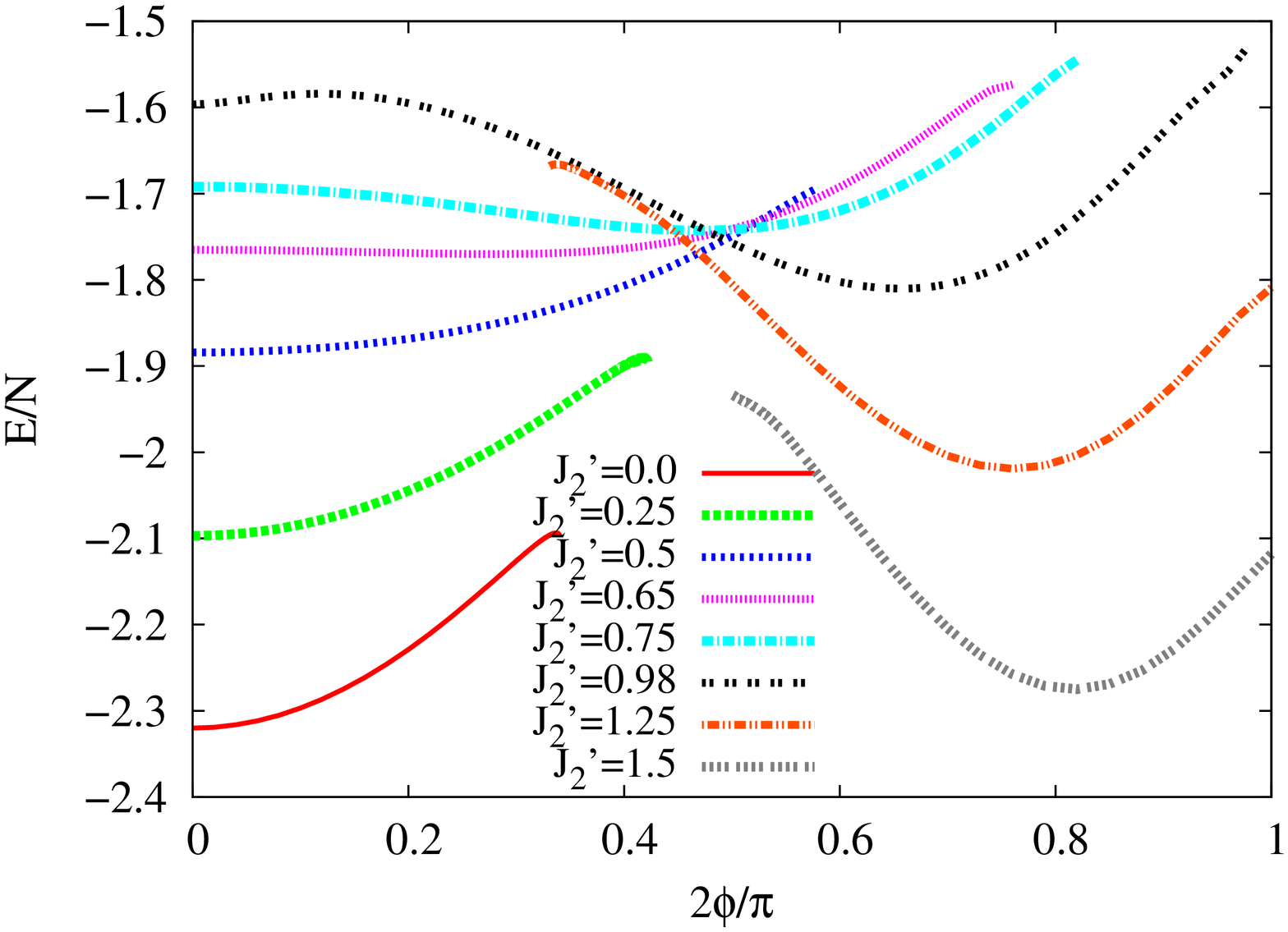}}}
\subfloat[$s=1$]{\scalebox{0.3}{\includegraphics[angle=270]{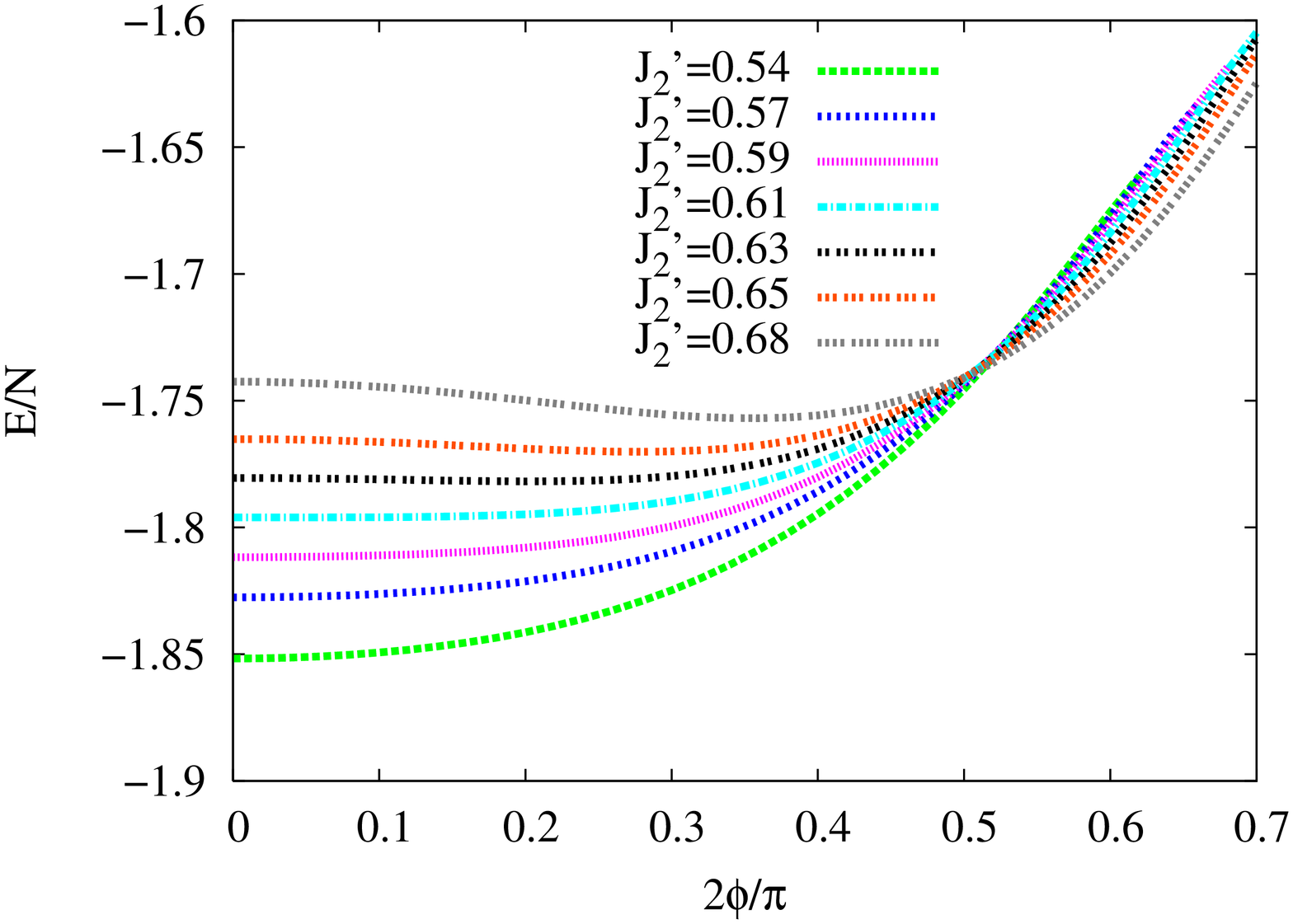}}}
}
\mbox{
 \subfloat[$s=\frac{3}{2}$]{\scalebox{0.3}{\includegraphics[angle=270]{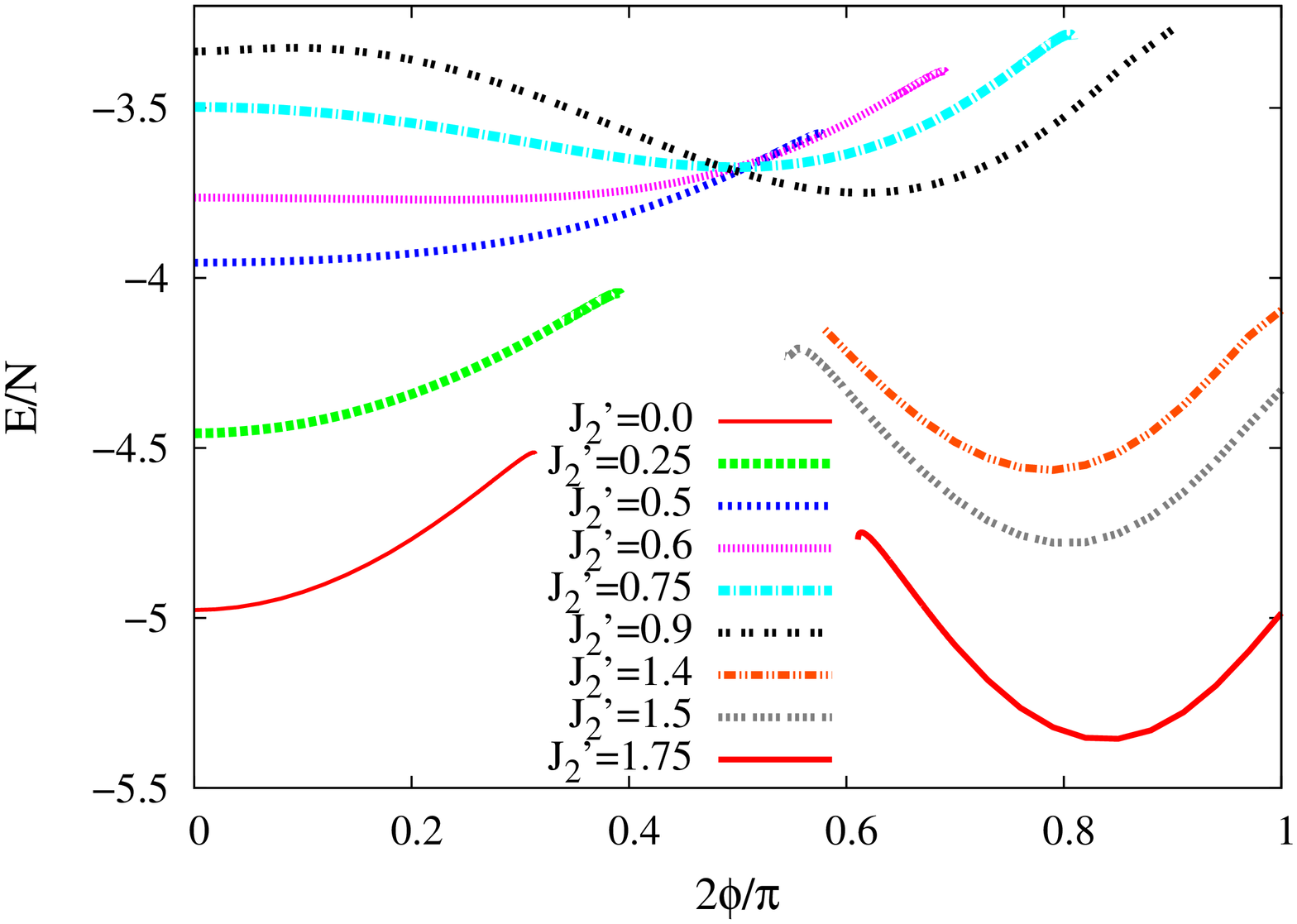}}}
\subfloat[$s=\frac{3}{2}$]{\scalebox{0.3}{\includegraphics[angle=270]{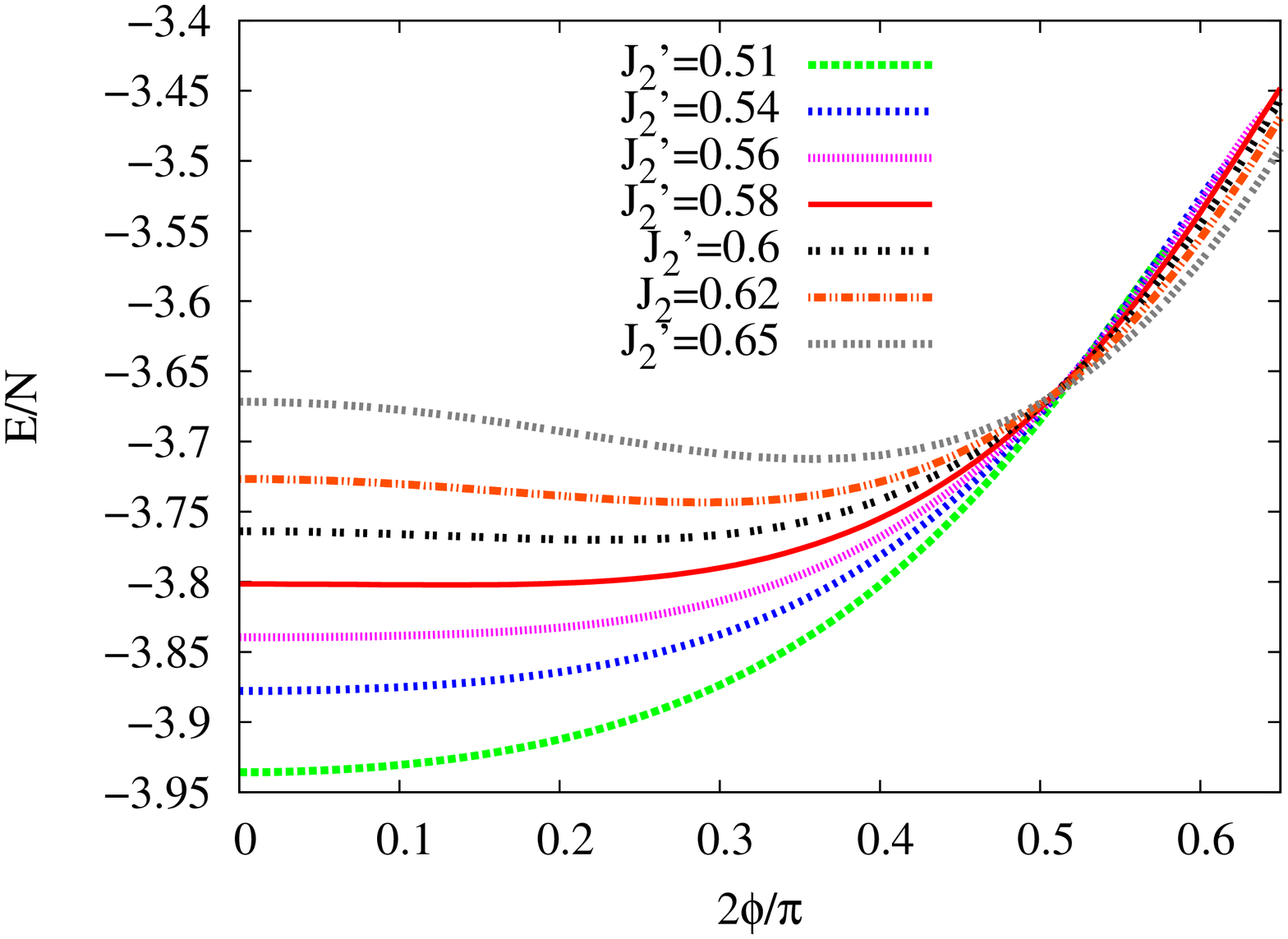}}}
}
\caption{(colour online) Ground-state energy per spin of the spin-1 and
  spin-$\frac{3}{2}$ $J_{1}$-$J_{2}'$ Hamiltonian of Eq.\ (\ref{H}) with
  $J_{1}=1$, using the SUB$4$-$4$ approximation of the CCM with the
  spiral model state, versus the spiral angle $\phi$. For the case of
  $s=1$, for $J_{2}' \lesssim 0.610$ the only minimum is at $\phi=0$
  (N\'{e}el order), whereas for $J_{2}' \gtrsim 0.610$ a secondary
  minimum occurs at $\phi=\phi_{{\rm SUB}4-4} \neq 0$, which is also a
  global minimum, thus illustrating the typical scenario of a
  second-order phase transition. Similarly, for the case of $s=\frac{3}{2}$, for
  $J_{2}' \lesssim 0.571$ the only minimum is at $\phi=0$ (N\'{e}el
  order), whereas for $J_{2}' \gtrsim 0.571$ a secondary minimum
  occurs at $\phi=\phi_{{\rm SUB}4-4} \neq 0$.}
\label{EvsAngle}
\end{center}
\end{figure*}

\begin{figure*}[p!]
\begin{center}
\mbox{
  \subfloat[$s=1$]{\scalebox{0.3}{\includegraphics[angle=270]{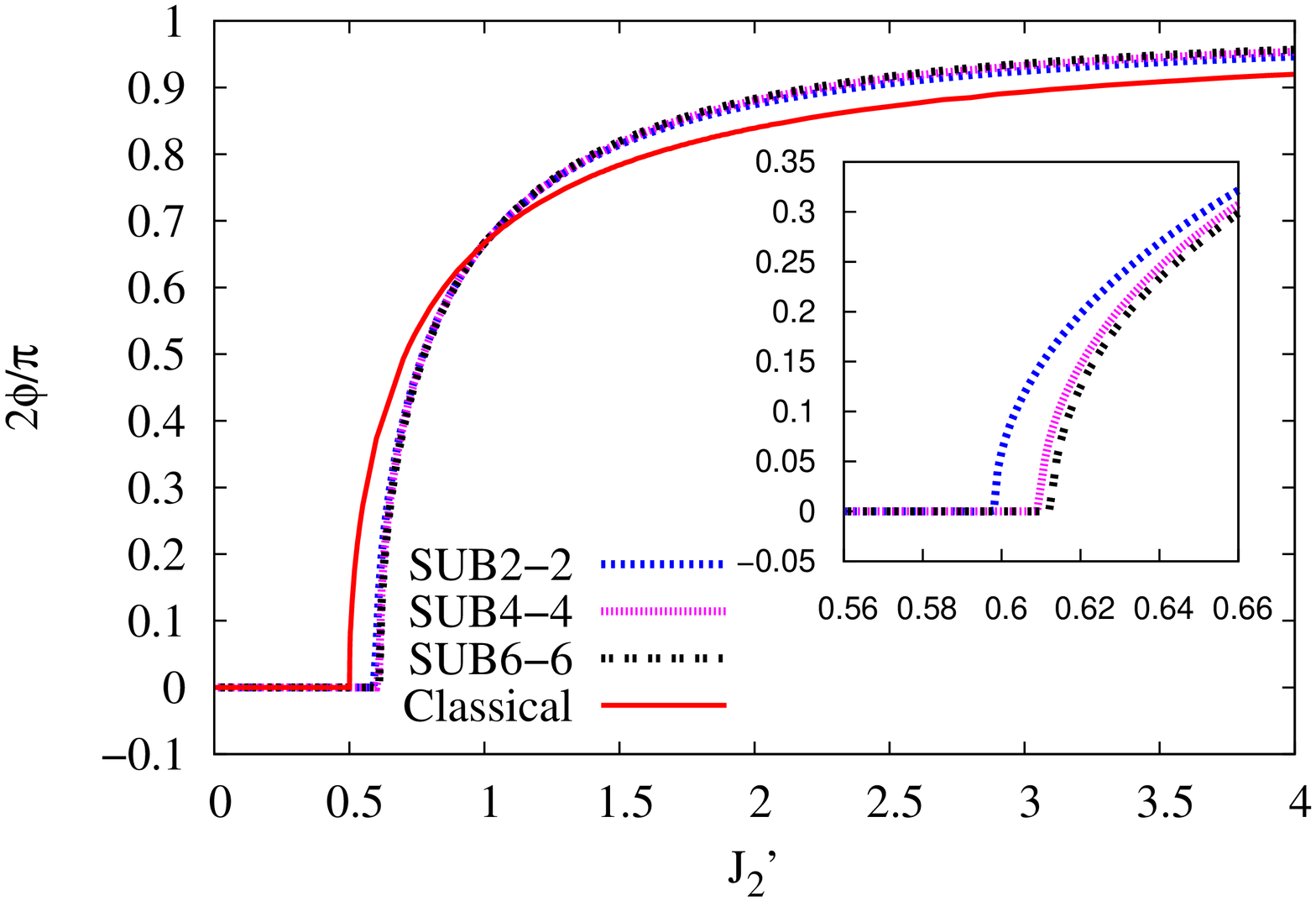}}}
  \subfloat[$s=1$]{\scalebox{0.3}{\includegraphics[angle=270]{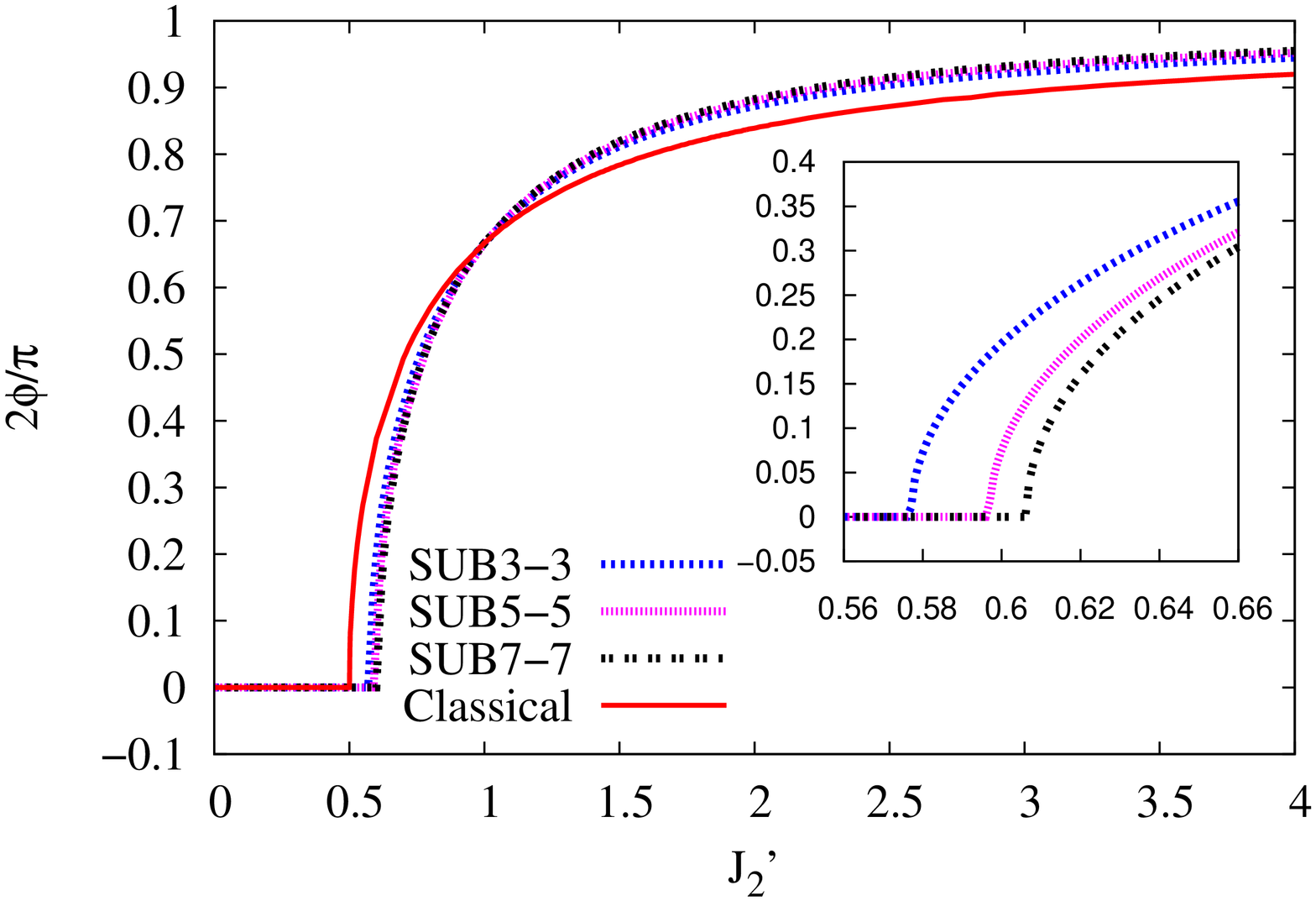}}}
}
\mbox{
 \subfloat[$s=\frac{3}{2}$]{\scalebox{0.3}{\includegraphics[angle=270]{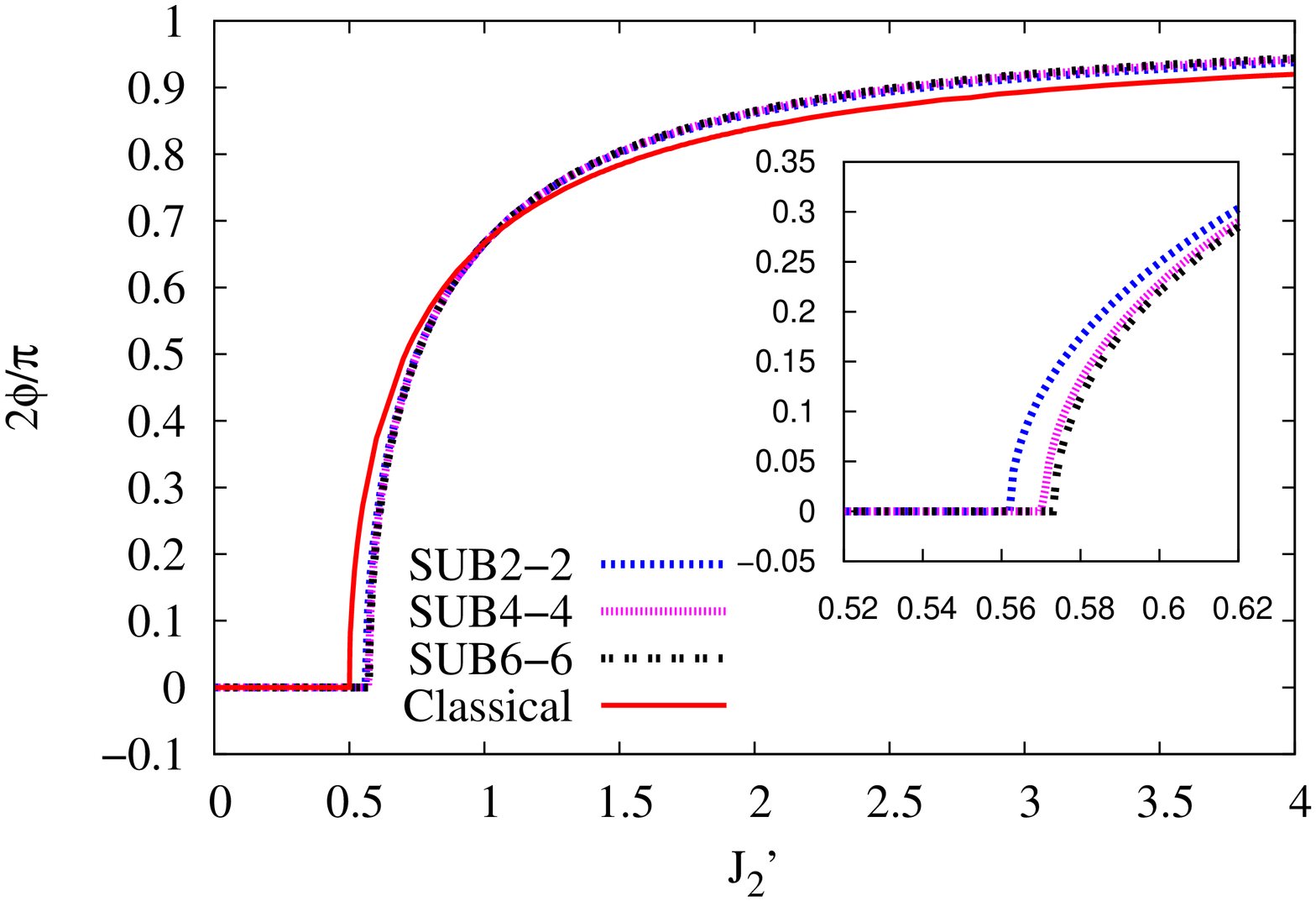}}}
 \subfloat[$s=\frac{3}{2}$]{\scalebox{0.3}{\includegraphics[angle=270]{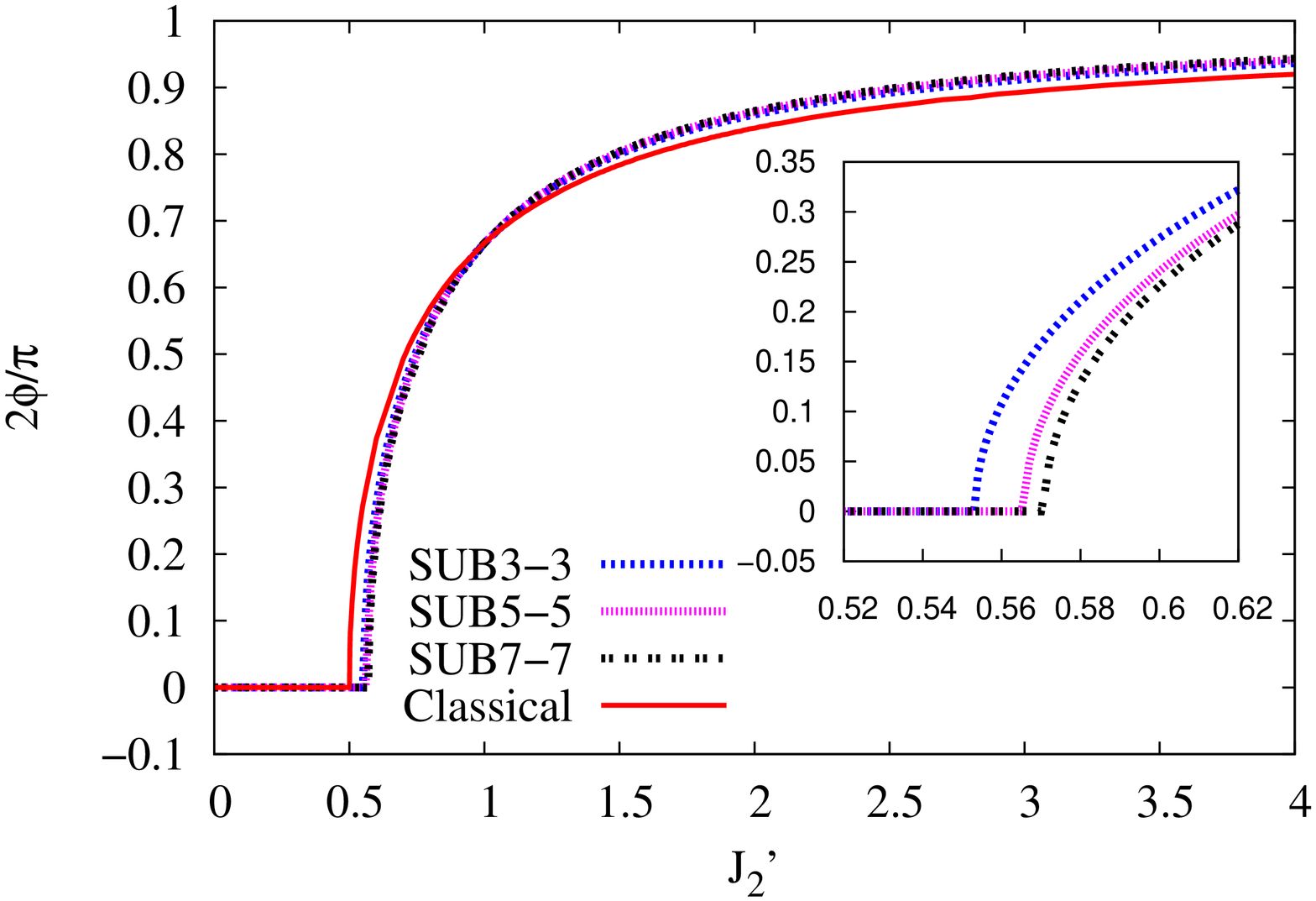}}}
}
\caption{(colour online) The angle $\phi_{{\rm SUB}n-n}$ that
  minimizes the energy $E_{{\rm SUB}n-n}(\phi)$ of the spin-1 and
  spin-$\frac{3}{2}$ $J_{1}$-$J_{2}'$ Hamiltonian of Eq.\ (\ref{H}) with
  $J_{1}=1$, in the SUB$n$-$n$ approximations with $n=\{2,4,6\}$ and
  $n=\{3,5,7\}$, using the spiral model state, versus $J_{2}'$.  The
  corresponding classical result $\phi_{{\rm cl}}$ is shown for
  comparison. We find in the SUB$n$-$n$ quantum case a
  second-order phase transition (e.g., at the SUB6-6 level, at
  $J_{2}' \approx 0.613$ for the $s=1$ case and at $J_{2}' \approx 0.574$
  for the $s=\frac{3}{2}$ case), where $\phi_{{\rm SUB}n-n}$ changes continuously
  from zero below the transition point (N\'{e}el phase) to a nonzero value
  above it (helical phase).  The classical case has a
  second-order phase transition at $J_{2}'=0.5$.}
\label{angleVSj2}
\end{center}
\end{figure*}

Classically we have a second-order phase transition from
N\'{e}el order (for $\kappa < \kappa_{{\rm cl}}$) to helical order
(for $\kappa > \kappa_{{\rm cl}}$), where $\kappa \equiv
J_{2}'/J_{1}$, at a value $\kappa_{{\rm cl}} = 0.5$.  By contrast, our CCM
results presented below show that there is a shift of this critical point to a value
$\kappa_{c} \approx 0.615 \pm 0.010$ in the spin-1 quantum case 
and $\kappa_{c} \approx 0.575 \pm 0.005$ for the spin-$\frac{3}{2}$ quantum case, 
first indications of which are seen in Figs.\ \ref{EvsAngle} and \ref{angleVSj2}.
%%%%%%%%%%%%%%%%%%%%
%%%%%%%%%%%%%%%%%%%%%%%
%%%%%%%%%%%%%%%%%
In both cases this is a second-order phase transition from 
N\'{e}el-ordered to helically-ordered states.
Thus, for example, curves such as those shown in Fig.\ \ref{EvsAngle}
show that the N\'{e}el state ($\phi=0$) gives the minimum gs
energy for all values of $\kappa < \kappa_{c}$, where
$\kappa_{c}$ depends on the level of SUB$n$-$n$ approximation
used, as we also observe in Fig.\ \ref{angleVSj2}. By contrast, for
values of $\kappa > \kappa_{c}$ the minimum in the energy is found
to occur at a value $\phi \neq 0$. If we consider the pitch angle
$\phi$ itself as an order parameter (i.e., $\phi=0$ for N\'{e}el order
and $\phi \neq 0$ for spiral order) a typical scenario for a phase
transition would be the appearance of a two-minimum structure for the
gs energy for values of $\kappa > \kappa_{c}$, exactly as observed
in Fig.\ \ref{EvsAngle} for both the spin-1 and spin-$\frac{3}{2}$ models in the
SUB4-4 approximation. Very similar curves occur for other
SUB$n$-$n$ approximations.

We note that the crossover from one minimum ($\phi=0$,
N\'{e}el) solution to the other ($\phi \neq 0$, spiral) appears to be
quite smooth at this point (and see Figs.\ \ref{EvsAngle} and
\ref{angleVSj2}). Thus, for example, the spiral pitch angle $\phi$ 
appears to change quite continuously from a value of zero for 
$\kappa < \kappa_{c}$ on the N\'{e}el side of the transition to a 
nonzero value for $\kappa > \kappa_{c}$ on the spiral-phase side.  For 
example, at the SUB$6$-$6$ level we find $\kappa_{c} \approx 0.613$ for 
the spin-1 case, and $\kappa_{c} \approx 0.574$ for 
the spin-spin-$\frac{3}{2}$ case.
We also note from Fig.~\ref{angleVSj2} that as $J_{2}
\rightarrow \infty$ the spiral angle $\phi$ approaches the limiting value 
$\frac{1}{2}\pi$ considerably slower for the spin-1 and spin-$\frac{3}{2}$ cases
than it does the spin-$\frac{1}{2}$ case we investigated earlier (and see Fig. 3 
in Ref. \cite{Bi:2008_SqTrian}).  This is a first indication that there 
is less freedom for the existence of a stable collinear (striped) state at 
higher values of $\kappa$ for the higher spin ($s > \frac{1}{2}$) models than 
for the $s =\frac{1}{2}$ model.  We return to this point later.  

Figure\ \ref{EvsAngle} shows the ground-state energy per spin versus the
spiral angle $\phi$, using the SUB$4$-$4$ approximation of the CCM
with the spiral model state, for some illustrative values of $J_{2}'$.
Similarly Fig.\ \ref{angleVSj2} shows the angle $\phi_{{\rm SUB}n-n}$ that
minimizes the energy $E_{{\rm SUB}n-n}(\phi)$. Our previous study of the
quantum spin-$\frac{1}{2}$ case in the same
model \cite{Bi:2008_SqTrian} found that there is a first
quantum critical point at $\kappa_{c_{1}} \approx 0.80$ 
at which a weakly first-order, or possibly second-order, phase 
transition occurs between states that exhibit N\'{e}el order and 
helical order.  We see now that increasing the spin 
quantum number $s$ thus brings the quantum
critical point $\kappa_{c}$ closer to the classical critical point
$\kappa_{{\rm cl}}=0.5$ for the phase transition from N\'{e}el order
to helical order, as expected.

We observe from Fig.\ \ref{EvsAngle} that for certain values of
$J_{2}'$ (or, equivalently, $\kappa$) CCM solutions at a given SUB$n$-$n$
level of approximation (viz., SUB$4$-$4$ in Fig.\ \ref{EvsAngle})
exist only for certain ranges of the spiral angle $\phi$.  For example,
for the pure square-lattice HAF ($\kappa=0$) the CCM SUB$4$-$4$
solution based on a spiral model state only exists for $0 \leq \phi
\lesssim 0.17 \pi$ for the spin-1 model and $0 \leq \phi
\lesssim 0.16 \pi$ for the spin-$\frac{3}{2}$ model. In this case, where the N\'{e}el
solution is the stable ground state, if we attempt to move too far
away from N\'{e}el collinearity the CCM equations themselves become
``unstable'' and simply do not have a real solution.  Similarly, we
see from Fig.\ \ref{EvsAngle} that for $\kappa=1.5$ the CCM
SUB$4$-$4$ solution exists only for $0.25 \pi \lesssim \phi \leq 0.5
\pi$ for the spin-1 model and for $0.27 \pi \lesssim \phi \leq 0.5
\pi$ for the spin-$\frac{3}{2}$ model. In this case the stable
ground state is a spiral phase, and now if we attempt to move too
close to N\'{e}el collinearity the real solution terminates.

Such terminations of CCM solutions are common \cite{Fa:2004}. A
termination point usually arises because the solutions to the CCM equations become
complex at this point, beyond which there exist two branches of
entirely unphysical complex conjugate solutions \cite{Fa:2004}. In the
region where the solution reflecting the true physical solution is
real there actually also exists another (unstable) real
solution. However, only the (shown) upper branch of these two
solutions reflects the true (stable) physical ground state, whereas
the lower branch does not. The physical branch is usually easily
identified in practice as the one which becomes exact in some known
(e.g., perturbative) limit. This physical branch then meets (with
infinite slope, as seen in Fig.\ \ref{EvsAngle}) the corresponding unphysical 
branch at some termination point beyond which no real solutions
exist. The SUB$n$-$n$ termination points are themselves also
reflections of the quantum phase transitions in the real system, and
may be used to estimate the position of the phase
boundary \cite{Fa:2004}, although we do not do so here 
since we have more accurate criteria discussed below.

\begin{figure*}[t!]
\begin{center}
\mbox{
  \subfloat[$s=1$]{\scalebox{0.3}{\includegraphics[angle=270]{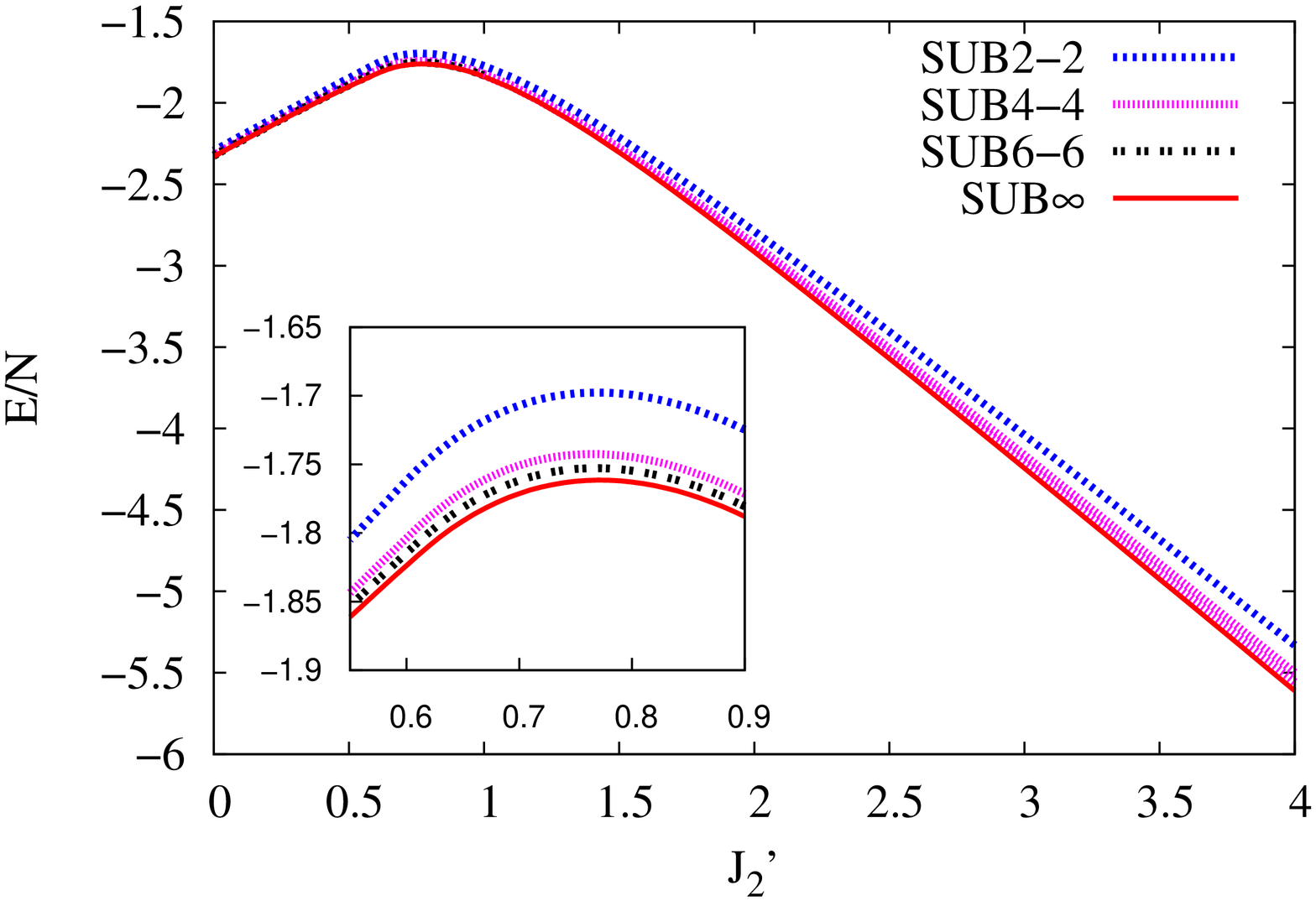}}}
\subfloat[$s=1$]{\scalebox{0.3}{\includegraphics[angle=270]{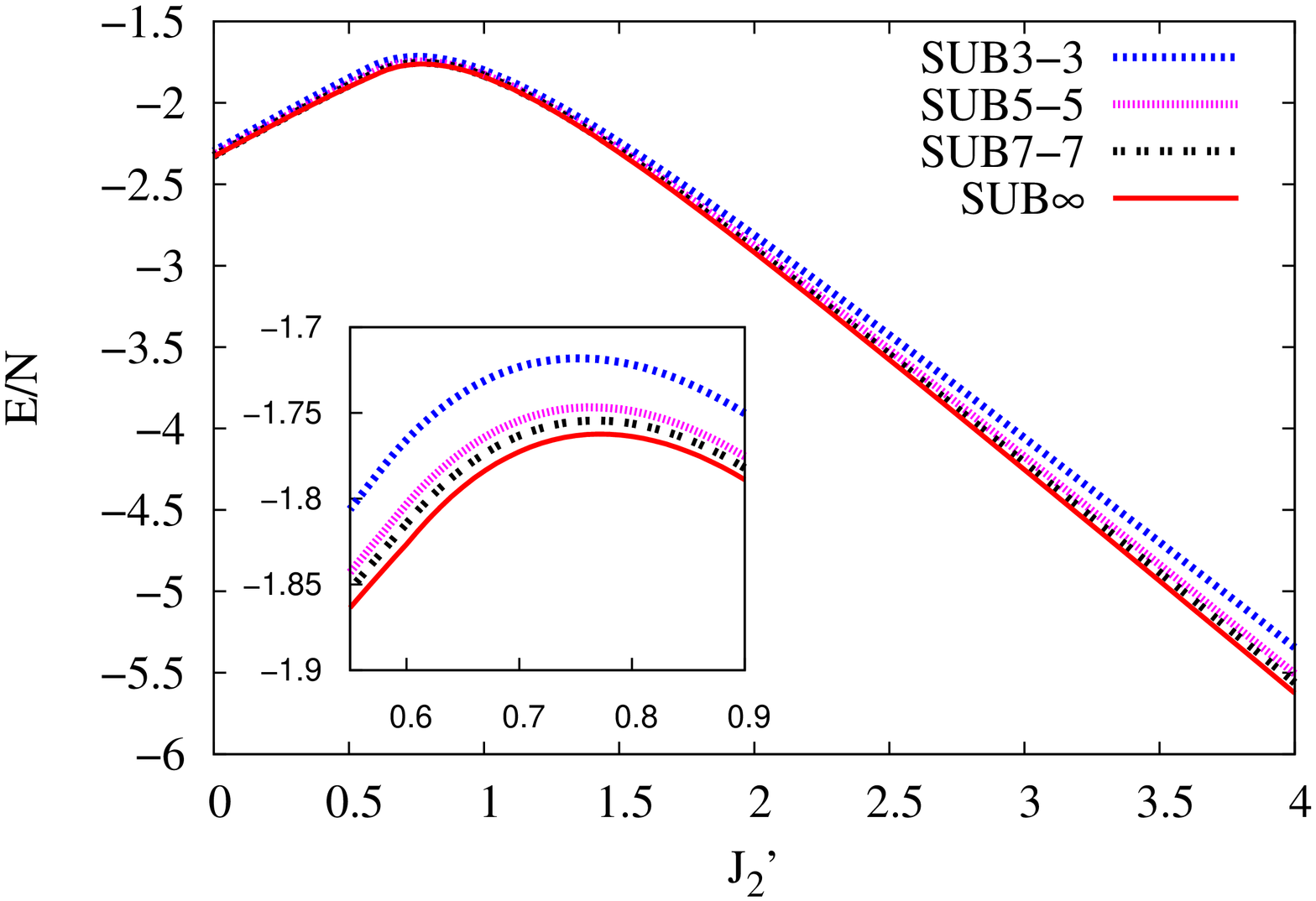}}}
}
\mbox{
 \subfloat[$s=\frac{3}{2}$]{\scalebox{0.3}{\includegraphics[angle=270]{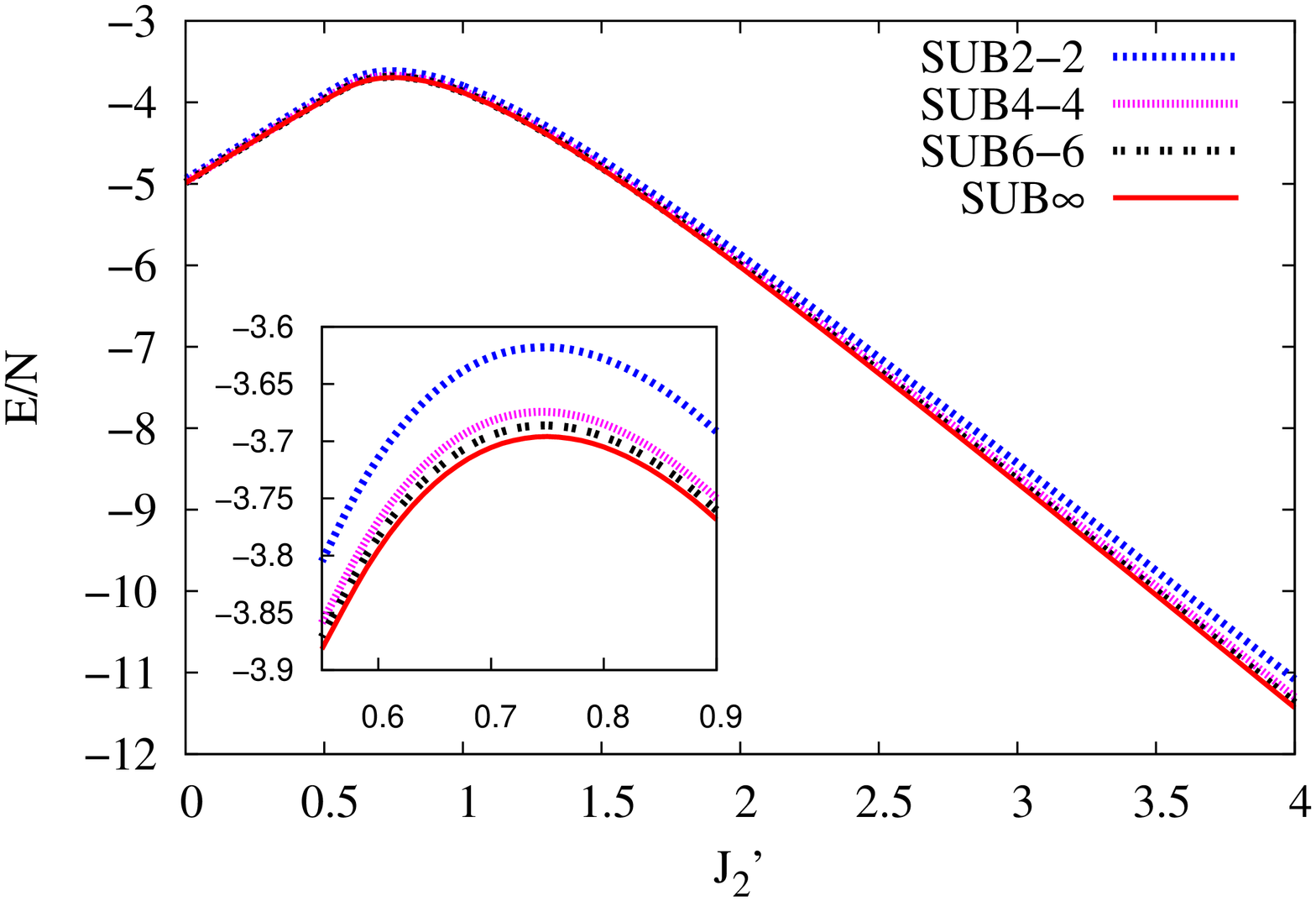}}}
  \subfloat[$s=\frac{3}{2}$]{\scalebox{0.3}{\includegraphics[angle=270]{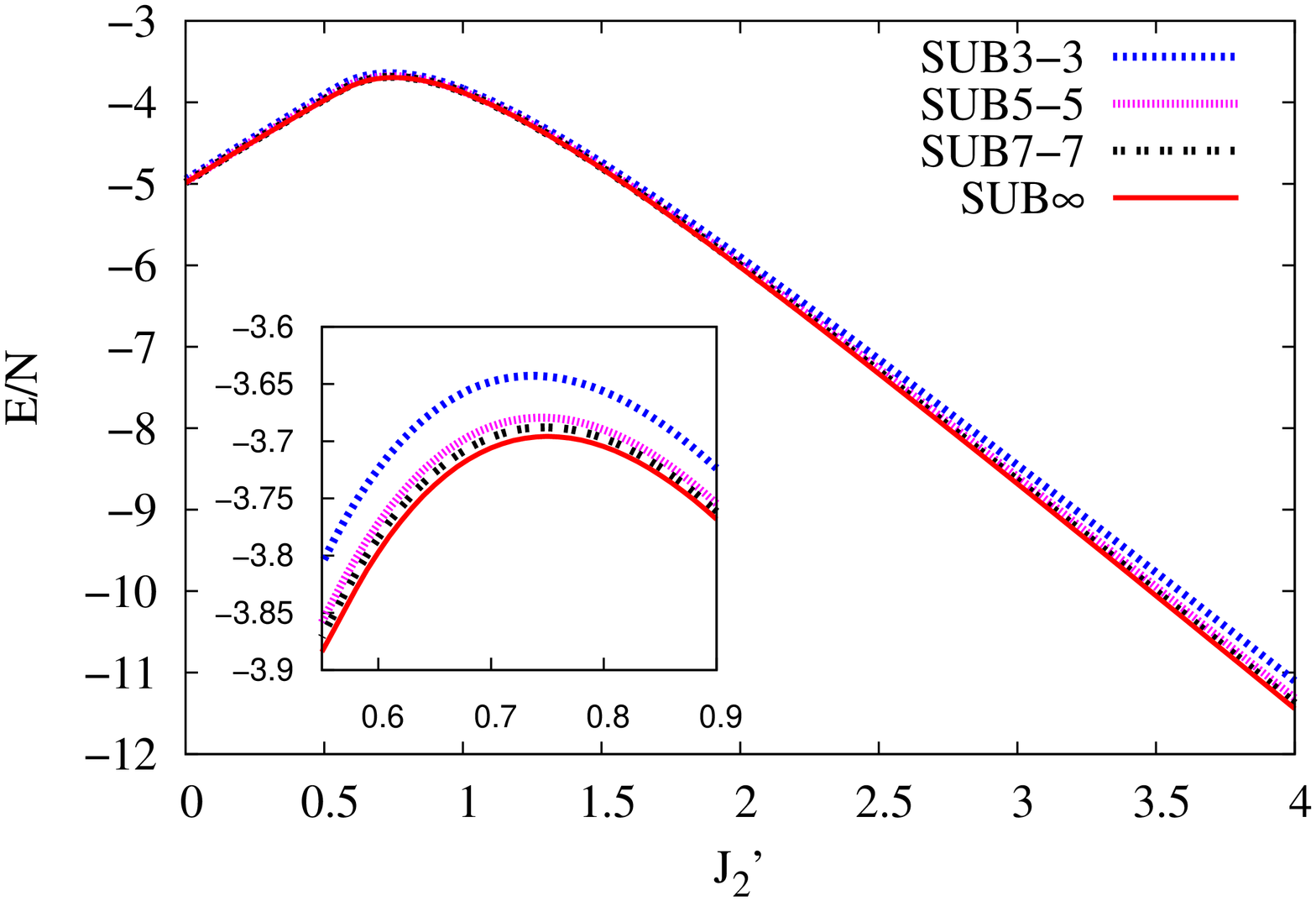}}}
}
\caption{(colour online) Ground-state energy per spin versus $J_{2}'$
  for the N\'{e}el and spiral phases of the spin-1 and spin-$\frac{3}{2}$ $J_{1}$-$J_{2}'$ Hamiltonian of Eq.\ (\ref{H})
  with $J_{1}=1$. The CCM results
  using the spiral model state are shown for various SUB$n$-$n$
  approximations ($n=\{2,4,6\}$) and ($n=\{3,5,7\}$) with the spiral angle
  $\phi=\phi_{{\rm SUB}n-n}$ that minimizes $E_{{\rm SUB}n-n}(\phi)$.
  We also show the $n \rightarrow \infty$ extrapolated results from
  using Eq.\ (\ref{Extrapo_E}).}
\label{E}
\end{center}
\end{figure*}

\begin{figure*}[p!]
\begin{center}
\mbox{
\subfloat[$s=1$]{\scalebox{0.3}{\includegraphics[angle=270]{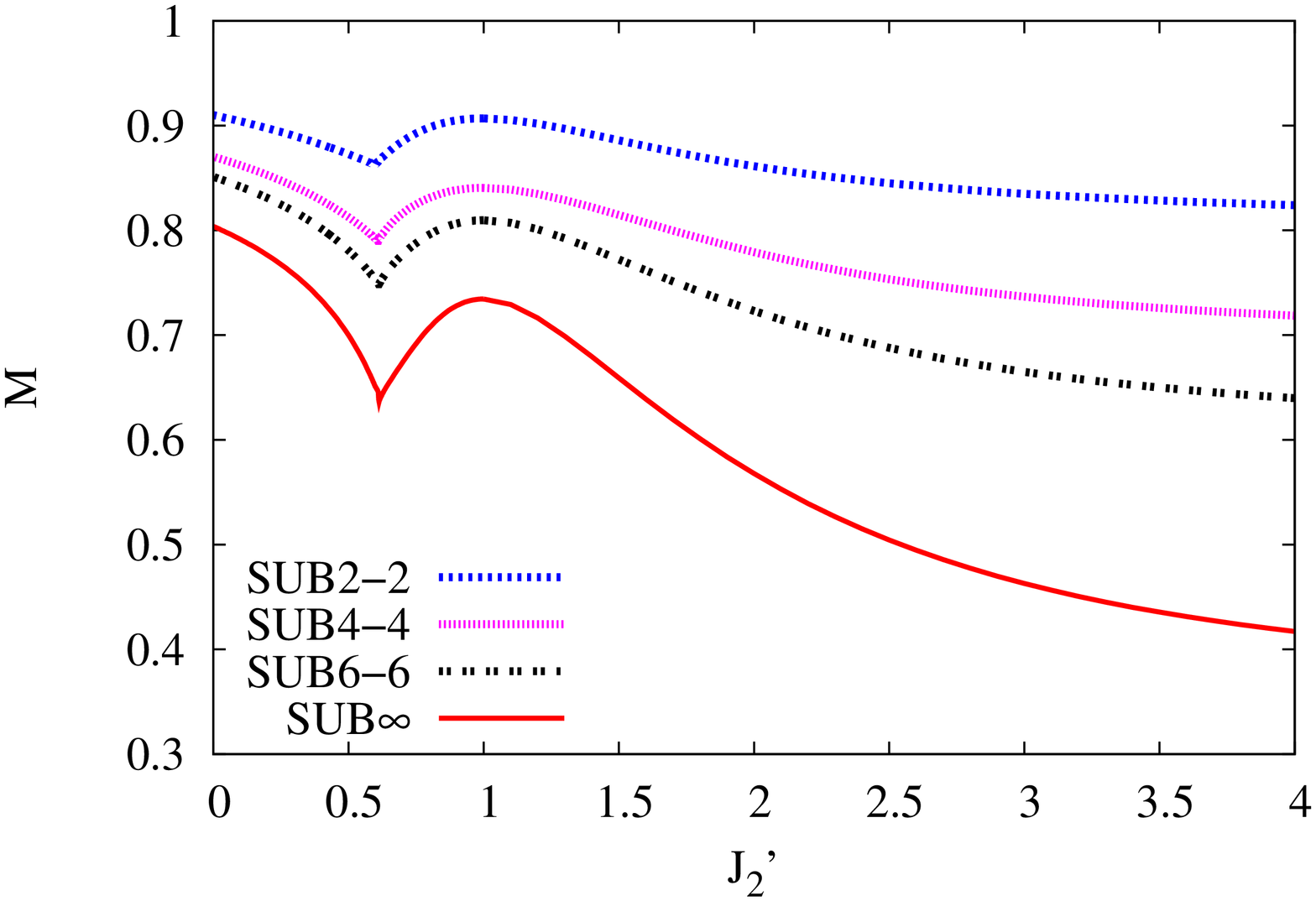}}}
\subfloat[$s=1$]{\scalebox{0.3}{\includegraphics[angle=270]{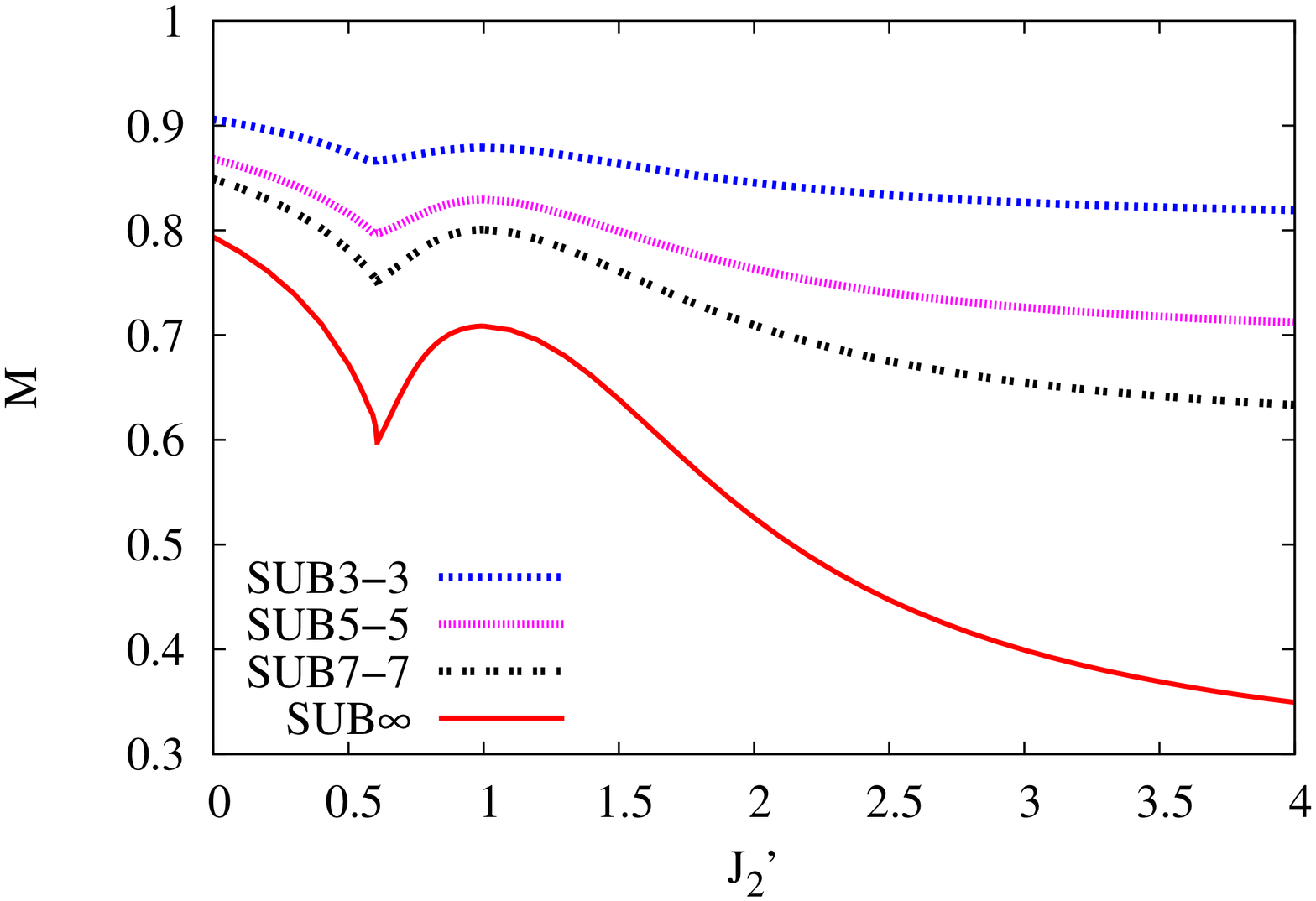}}}
%\subfloat[$s=\frac{3}{2}$]{\scalebox{0.3}{\includegraphics[angle=270]{fig5b.eps}}}
}
\mbox{
%\subfloat[$s=1$]{\scalebox{0.3}{\includegraphics[angle=270]{fig5c.eps}}}
\subfloat[$s=\frac{3}{2}$]{\scalebox{0.3}{\includegraphics[angle=270]{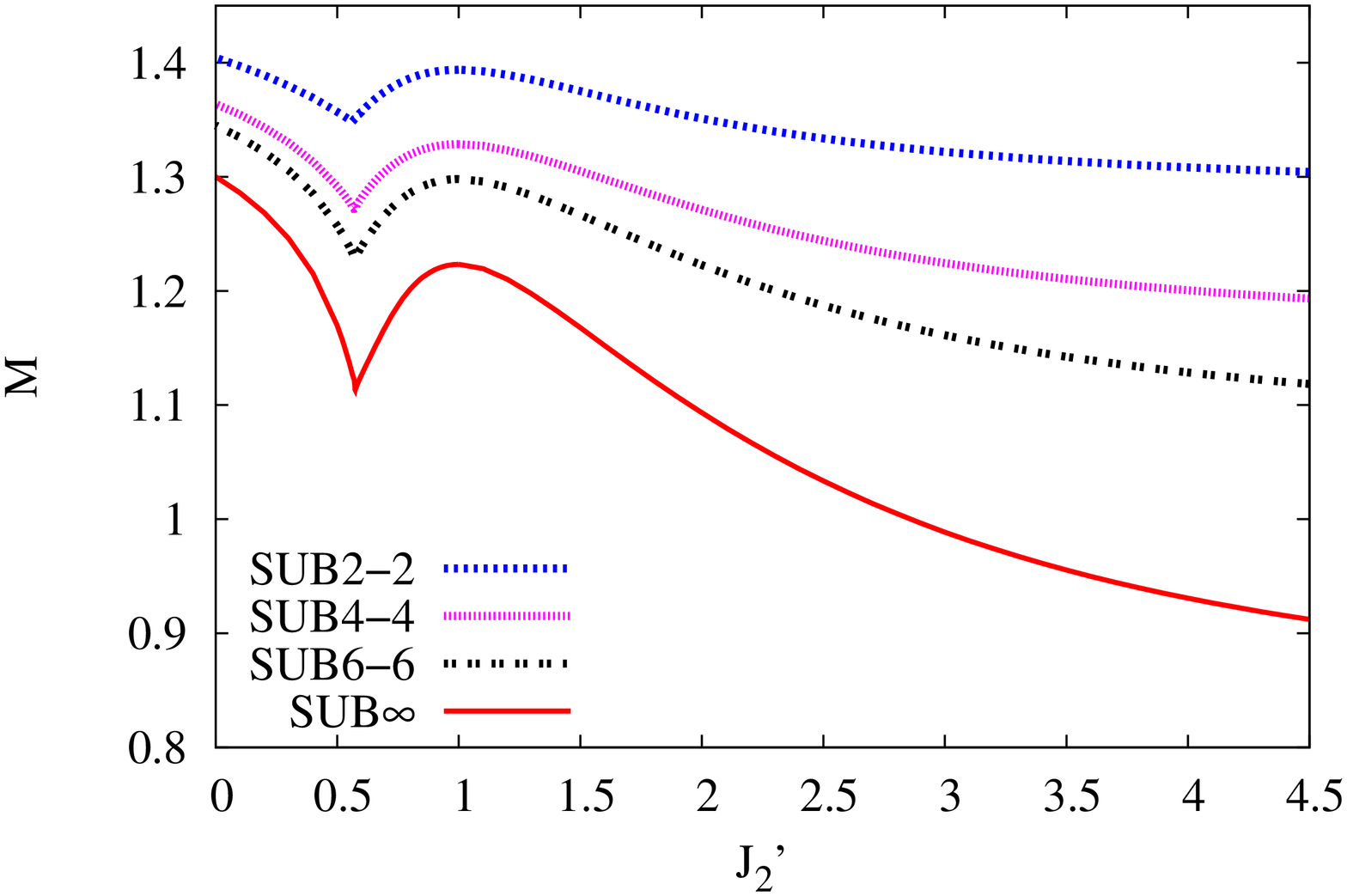}}}
\subfloat[$s=\frac{3}{2}$]{\scalebox{0.3}{\includegraphics[angle=270]{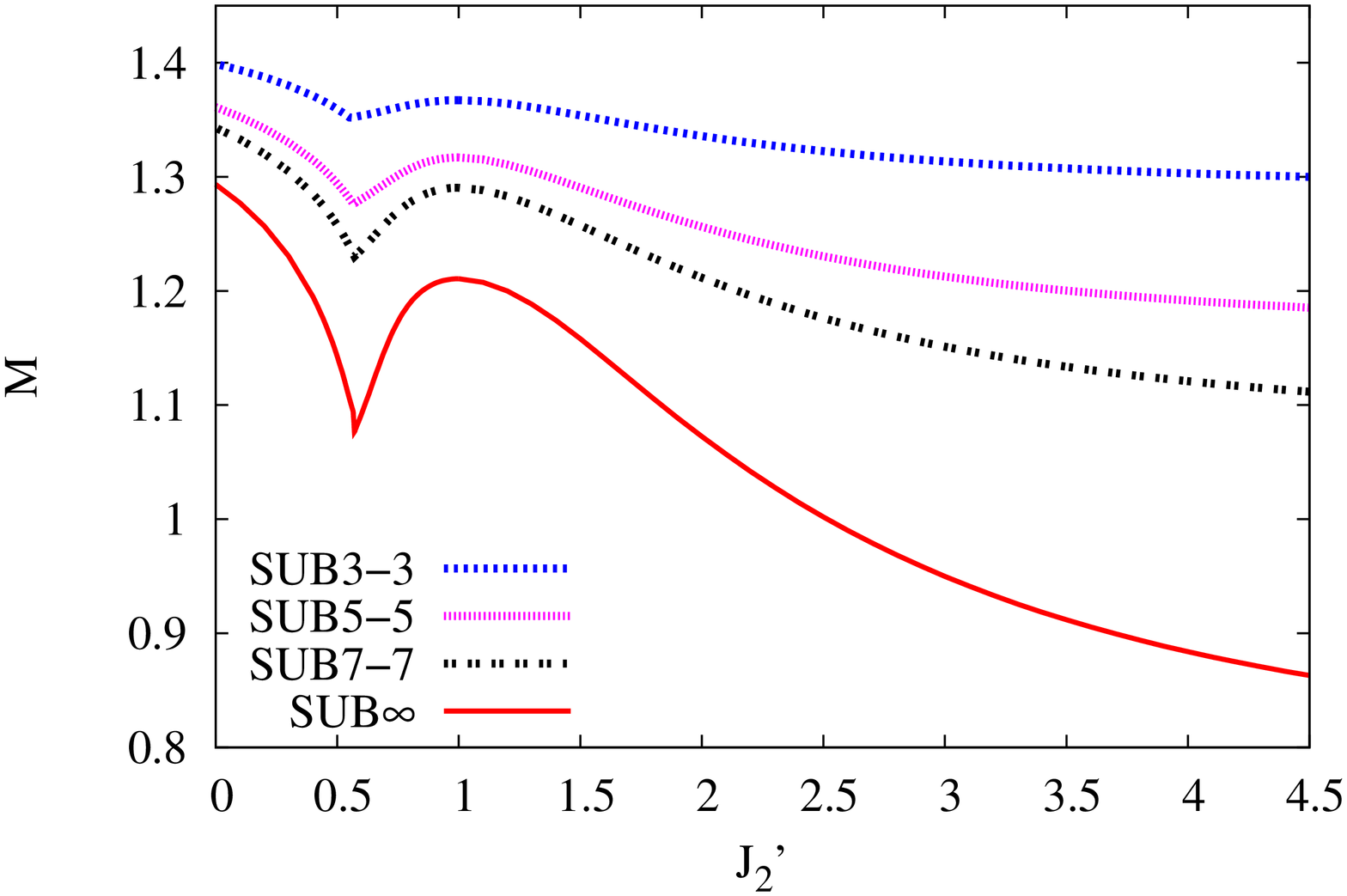}}}
}
\caption{(colour online) Ground-state magnetic order parameter (i.e.,
  the average on-site magnetization) versus $J_{2}'$ for
  the N\'{e}el and spiral phases of the spin-1 and spin-$\frac{3}{2}$
  $J_{1}$-$J_{2}'$ Hamiltonian of Eq.\ (\ref{H}) with $J_{1}=1$. The CCM results using the spiral
  model state are shown for various SUB$n$-$n$ approximations
  ($n=\{2,4,6\}$ and ($n=\{3,5,6\}$) with the spiral angle $\phi=\phi_{{\rm SUB}n-n}$
  that minimizes $E_{{\rm SUB}n-n}(\phi)$. We also show the $n
  \rightarrow \infty$ extrapolated results from using Eq.\
  (\ref{Extrapo_M}). The sharp minimum in the extrapolated magnetic order parameter 
  is at $J_{2}'=0.613$ ($M=0.6367$) using $n=\{2,4,6\}$ and $J_{2}'=0.606$ ($M = 0.5978$) using 
  $n=\{3,5,7\}$ for the spin-1 case, whereas for the spin-$\frac{3}{2}$ case, the 
  corresponding values are $J_{2}'=0.574$ ($M = 1.1134$) using 
  $n=\{2,4,6\}$ and $J_{2}'=0.571$ ($M = 1.0766$) using $n=\{3,5,7\}$.}
\label{M}
\end{center}
\end{figure*}

Figures \ref{E} and \ref{M} 
%%%%%%%%%%%%%%%%%%%
show the CCM results for the gs energy and average gs on-site magnetization,
respectively, where the spiral state has been used as the model
state. The gs energy (in Fig.\ \ref{E}) shows no sign of a discontinuity
in slope at the critical values $\kappa_{c}$ discussed above, and this
is an indication of a second-order transition from the N\'{e}el phase
to the helical phase. This is in contrast with the spin-$\frac{1}{2}$ case,
where the gs energy shows definite signs of a (weak) discontinuity in
slope at the first critical
value $\kappa_{c_{1}}$ \cite{Bi:2008_SqTrian}.

The gs magnetic order parameter $M$ in Fig.\ \ref{M} shows much clearer 
evidence of a phase transition at the
corresponding $\kappa_{c}$ values previously observed in Fig.\
\ref{angleVSj2}.  Thus, we see that for the spin-1 case the sharp minimum in the extrapolated 
magnetic order parameter occurs at $\kappa_{c} \approx 0.613$ (with $M_{c}=0.6367$) 
using $n=\{2,4,6\}$ for the extrapolation of $M$, and at $\kappa_{c} \approx 0.606$ 
(with $M_{c} = 0.5978$) using $n=\{3,5,7\}$; 
whereas for the spin-$\frac{3}{2}$ case, the corresponding values are 
$\kappa_{c} \approx 0.574$ (with $M_{c} = 1.1134$) using $n=\{2,4,6\}$ for the 
extrapolation of $M$ and at $\kappa_{c} \approx 0.571$ (with $M_{c} = 1.0766$) 
using $n=\{3,5,7\}$.  We also present other independent estimates 
for $\kappa_{c}$ below. 

By contrast, for the spin-$\frac{1}{2}$ case \cite{Bi:2008_SqTrian} 
the extrapolated value of $M$ showed clearly its steep
drop toward a value very close to zero at a corresponding value 
$\kappa_{c} \approx 0.80$,  
which gave the best CCM estimate of the phase-transition
point for that case.  In the spin-$\frac{1}{2}$ case the 
magnetization seemed to approach continuously a value  
$M=0.025 \pm 0.025$ from the N\'{e}el side ($\kappa < \kappa_{c}$)
whereas from the spiral side ($\kappa > \kappa_{c}$) there appeared 
to be a discontinuous jump in $M$ as $\kappa \rightarrow \kappa_{c}$.
The transition at $\kappa < \kappa_{c}$
thus appeared to be (very) weakly first order but it was not possible
to exclude it being second order since 
the possibility of a continuous but very steep drop to zero
of the on-site magnetization as $\kappa \rightarrow \kappa_{c}$ from 
the spiral side of the transition could not be entirely ruled out. No 
evidence at all was found for any intermediate phase between the quasiclassical
N\'{e}el and spiral phases, just as for the higher-spin cases considered here.  
However, Fig.\ \ref{M} here shows no evidence at all for a finite jump 
in $M$ as $\kappa \rightarrow \kappa_{c}$ from either side of the 
transition, and hence the evidence from the order parameter is that the 
transition from N\'{e}el order to spiral order for both the spin-1 
and spin-$\frac{3}{2}$ cases is of second-order type.

Table~\ref{table_CritPt}
\begin{table}[tb!]
%\label{table_CritPt}
\caption{The critical value $\kappa_{c} = \kappa^{{\rm SUB}n-n}_{c}$ at which the 
transition between the N\'{e}el phase ($\phi=0$) and the spiral phase 
($\phi \neq 0$) occurs in various SUB$n$-$n$ approximations, using the CCM 
with the (N\'{e}el or) spiral state as model state, for the $J_{1}$-$J_{2}'$ 
model.  Results are shown for both the spin-1 and spin-$\frac{3}{2}$ cases.}
\vskip0.5cm
\begin{tabular}{cccc} 
\hline\noalign{\smallskip}
\multirow{2}{*}{Method} & $s=1$ & &  $s=\frac{3}{2}$ \\ 
\noalign{\smallskip}\cline{2-2} \cline{4-4}\noalign{\smallskip}
 & $\kappa_{c}$ & & $\kappa_{c}$ \\ 
\noalign{\smallskip}\hline\noalign{\smallskip}
SUB$2$-$2$ & 0.597 &  &   0.563   \\ 
SUB$4$-$4$ & 0.610 &   & 0.571   \\ 
SUB$6$-$6$ & 0.613  &  &  0.574  \\ 
SUB$\infty$ $^{a}$ & 0.617  &   & 0.581  \\ 
SUB$\infty$ $^{b}$ & 0.616 &   & 0.577  \\ \hline
SUB$3$-$3$ &  0.577  &  &    0.554  \\ 
SUB$5$-$5$ &  0.597  &     & 0.566 \\ 
SUB$7$-$7$ &    0.607 &     & 0.571 \\ 
SUB$\infty$ $^{c}$ &   0.636  & & 0.583 \\ 
SUB$\infty$ $^{d}$ &   0.619 &  & 0.577 \\ 
\noalign{\smallskip}\hline
\end{tabular} 
\\
$^{a}$ Based on $1/n$ : $n=\{2,4,6\}$ \\
$^{b}$ Based on $1/n^{2}$ : $n=\{2,4,6\}$ \\
$^{c}$ Based on $1/n$ : $n=\{3,5,7\}$ \\
$^{d}$ Based on $1/n^{2}$ : $n=\{3,5,7\}$ \\ 
\label{table_CritPt}
\end{table}
shows the critical values $\kappa^{{\rm SUB}n-n}_{c}$ at which the
transition between the N\'{e}el and spiral phases occurs in the
various SUB$n$-$n$ approximations shown in Fig.~\ref{angleVSj2}. In
the past we have found that a simple linear extrapolation scheme 
\cite{ccm_UJack_asUJ_2010,Bishop:2010_UJack_GrtSpins,Bishop:2010_KagomeSq},
$\kappa^{{\rm SUB}n-n}_{c} = a_{0}+a_{1}n^{-1}$, yields a good fit to
such critical points. This seems to be the case here too, just as for
the spin-$\frac{1}{2}$ case~\cite{Bi:2008_SqTrian}.  
The fact that the two corresponding ``SUB$\infty$'' estimates from the 
SUB$n$-$n$ data in Table~\ref{table_CritPt} based on the 
even-$n$ and odd-$n$ SUB$n$-$n$ sequences differ slightly from one 
another is a reflection of the errors inherent in our extrapolation 
procedures.  Similar estimates based on an alternative extrapolation 
scheme, $\kappa^{{\rm LSUB}n}_{c}=b_{0}+b_{1}n^{-2}$, are
also shown in Table~\ref{table_CritPt}.  The difference between all of these 
estimates is thus also a rough indication of our real error bars on 
$\kappa_{c}$.

It is gratifying to note that all of the estimates for $\kappa_{c}$ from
the extrapolations of our computed results for $\kappa^{{\rm SUB}n-n}_{c}$ 
are in excellent agreement with those obtained from the extrapolated 
results for the order parameter $M$ discussed above. By putting all of 
these results together, our final estimates for the critical point for the 
transition between the N\'{e}el-ordered and the spirally-ordered 
phases are $\kappa_{c}=0.615 \pm 0.010$ for the spin-1 model and
$\kappa_{c}=0.575 \pm 0.005$ for the spin-$\frac{3}{2}$ model.

We conclude our discussion of the N\'{e}el and spiral phases by 
presenting detailed results for the two spin cases for the two special limits 
of the model, namely the pure isotropic HAF on the square and triangular 
lattices.  Thus, Table~\ref{EandM_spiral}    
\begin{table*}[tbp!]
  \caption{Ground-state energy per spin and magnetic order parameter 
(i.e., the average on-site magnetization) for the spin-1 and spin-$\frac{3}{2}$ HAFs on 
the square and triangular lattices. We show CCM results obtained for the $J_{1}$-$J_{2}'$ model 
with $J_{1}>0$, using the spiral model state in various SUB$n$-$n$ approximations defined on 
the triangular lattice geometry, for the two cases $\kappa \equiv J_{2}'/J_{1}=0$ 
(square lattice HAF, $\phi=0$) and $\kappa=1$ (triangular lattice HAF, $\phi=\frac{\pi}{3}$).}
{\smallskip}
\label{EandM_spiral}
\begin{tabular}{ccccccccccccc}  
\hline\hline\noalign{\smallskip}
\multirow{3}{*}{Method} & \multicolumn{5}{c}{$s=1$} &  & \multicolumn{5}{c}{$s=\frac{3}{2}$} \\ 
\noalign{\smallskip}\cline{2-6}  \cline{8-12} \noalign{\smallskip}
&  {$E/N$} & {$M$}  & & {$E/N$} & {$M$} & & {$E/N$} & {$M$}  &  & {$E/N$} & {$M$} \\ 
\noalign{\smallskip} \cline{2-3} \cline{5-6}  \cline{8-9}   \cline{11-12} \noalign{\smallskip}
&   \multicolumn{2}{c}{square ($\kappa=0$)}  & & \multicolumn{2}{c}{triangular ($\kappa=1$)} & & \multicolumn{2}{c}{square ($\kappa=0$)} & & \multicolumn{2}{c}{triangular ($\kappa=1$)} \\  
\noalign{\smallskip} \hline \noalign{\smallskip}
SUB$2$-$2$ & -2.29504 & 0.9100 & & -1.77400 & 0.9069  & & -4.94393 & 1.4043 & & -3.80006 & 1.3938 \\ 
SUB$3$-$3$ & -2.29763 & 0.9059 & & -1.80101 & 0.8791  & & -4.94836 & 1.3990 & & -3.83393 & 1.3672 \\      
SUB$4$-$4$ & -2.31998 & 0.8702 & & -1.82231 & 0.8405  & & -4.97694 & 1.3638 & & -3.86025 & 1.3287 \\    
SUB$5$-$5$ & -2.32049 & 0.8682 & & -1.82623 & 0.8294  & & -4.97789 & 1.3611 & & -3.86498 & 1.3170 \\      
SUB$6$-$6$ & -2.32507 & 0.8510 & & -1.83135 & 0.8096  & & -4.98305 & 1.3452 & & -3.87059 & 1.2980\\             
SUB$7$-$7$ & -2.32535 & 0.8492 & & -1.83288 & 0.8006  & & -4.98344 & 1.3430 & & -3.87191 & 1.2904 \\     
\noalign{\smallskip} \hline \noalign{\smallskip}       
\multicolumn{12}{c}{Extrapolations} \\
\noalign{\smallskip} \hline \noalign{\smallskip}
SUB$\infty$ $^{a}$ & -2.32924 & 0.8038 & & -1.83860 & 0.7345 & & -4.98793  & 1.3001 & & -3.87869 & 1.2233  \\     
SUB$\infty$ $^{b}$ & -2.32975 & 0.7938 & & -1.83968 & 0.7086 & & -4.98803 & 1.2933 & & -3.87839  & 1.2107 \\ 
\noalign{\smallskip} \hline\hline \noalign{\smallskip}
CCM $^{c}$ & -2.3291 & 0.8067 & & & & & & & & & \\
SWT $^{d}$ & -2.3282 & 0.8043 & & & & & & & & & \\ 
SE $^{e}$ & -2.3279(2) & 0.8039(4) & & & & & & & & & \\
\noalign{\smallskip} \hline\hline \noalign{\smallskip}
\end{tabular}             
%\begin{flushleft}
\\ \protect $^{a}$ Based on $n=\{2,4,6\}$ \\ 
\protect $^{b}$ Based on $n=\{3,5,7\}$ \\ 
\protect $^{c}$ CCM (SUB$\infty$ for square lattice, based on $n=\{2,4,6\}$) in the natural square-lattice geometry \cite{Fa:2001_PRB64} \\ 
\protect $^{d}$ SWT (Spin-wave theory) for square lattice \cite{Ha:1992} \\
\protect $^{e}$ SE (Series Expansion) for square lattice \cite{Zh:1991}
%\end{flushleft} 
\end{table*}
%%%%%%%%%%%%%%%
shows the results for the ground-state energy per spin and magnetic order
parameter (i.e., the average on-site magnetization) for the spin-1 and
spin-$\frac{3}{2}$ $J_{1}$-$J_{2}'$ HAF model on the square lattice 
($J_{2}'=0$ or $\kappa =0$) and on the triangular lattice 
($J_{2}'=J_{1}$ or $\kappa = 1$), using the spiral
model state.  Our CCM results are presented in various SUB$n$-$n$
approximations (with $2 \leq n \leq 7$) based on the triangular
lattice geometry using the spiral model state, with $\phi=0$ for the
square lattice and $\phi=\frac{\pi}{3}$ for the triangular lattice.
The extrapolated results ($n \rightarrow \infty$) using Eqs.\
(\ref{Extrapo_E}) and (\ref{Extrapo_M}) with $n=\{2,4,6\}$ and
$n=\{3,5,7\}$ are also presented. For comparison we also 
show the results obtained for the spin-$1$ model on the square lattice 
(i.e., $\kappa = 0$) using spin-wave theory
(SWT) \cite{Ha:1992}, a linked-cluster series expansion
(SE) method \cite{Zh:1991}, and previous CCM SUB$n$-$n$ ($n \rightarrow
\infty$) results based on the model construed as referring to sites on a square 
lattice \cite{Fa:2001_PRB64}.  Our present results are seen both to 
be robust and internally consistent, by comparison of the independent
extrapolations of the SUB$n$-$n$ data using the even-$n$ and odd-$n$ 
data sets, and to agree very well with the best alternative results
available for the spin-$1$ model on the square lattice.  Such comparisons
give us confidence that our results are likely to be similarly accurate 
over the entire range of values of the frustration parameter $\kappa$.

\begin{figure*}[tbp!]
\begin{center}
\mbox{
\subfloat[$s=1$]{\scalebox{0.3}{\includegraphics[angle=270]{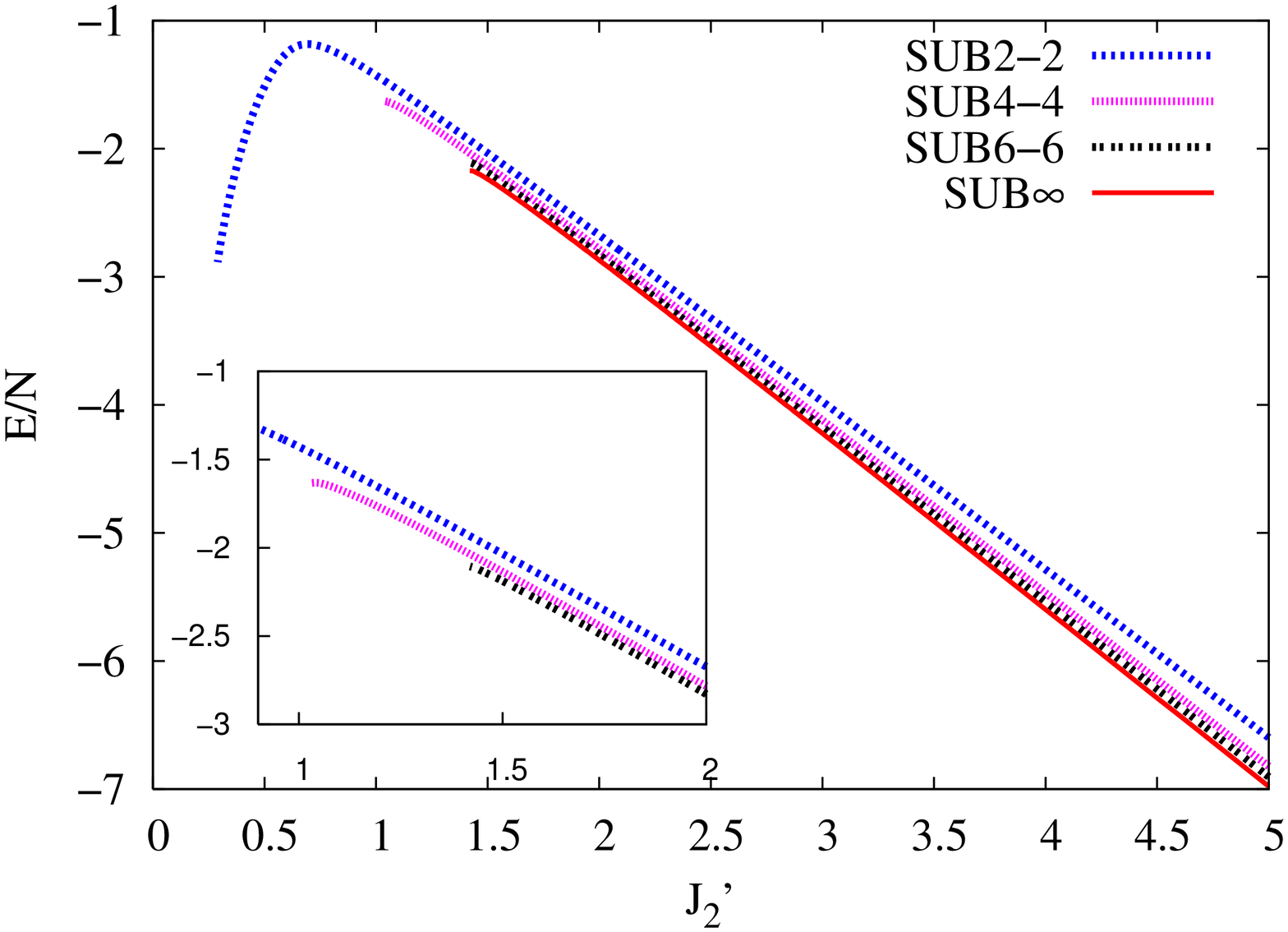}}}
\subfloat[$s=1$]{\scalebox{0.3}{\includegraphics[angle=270]{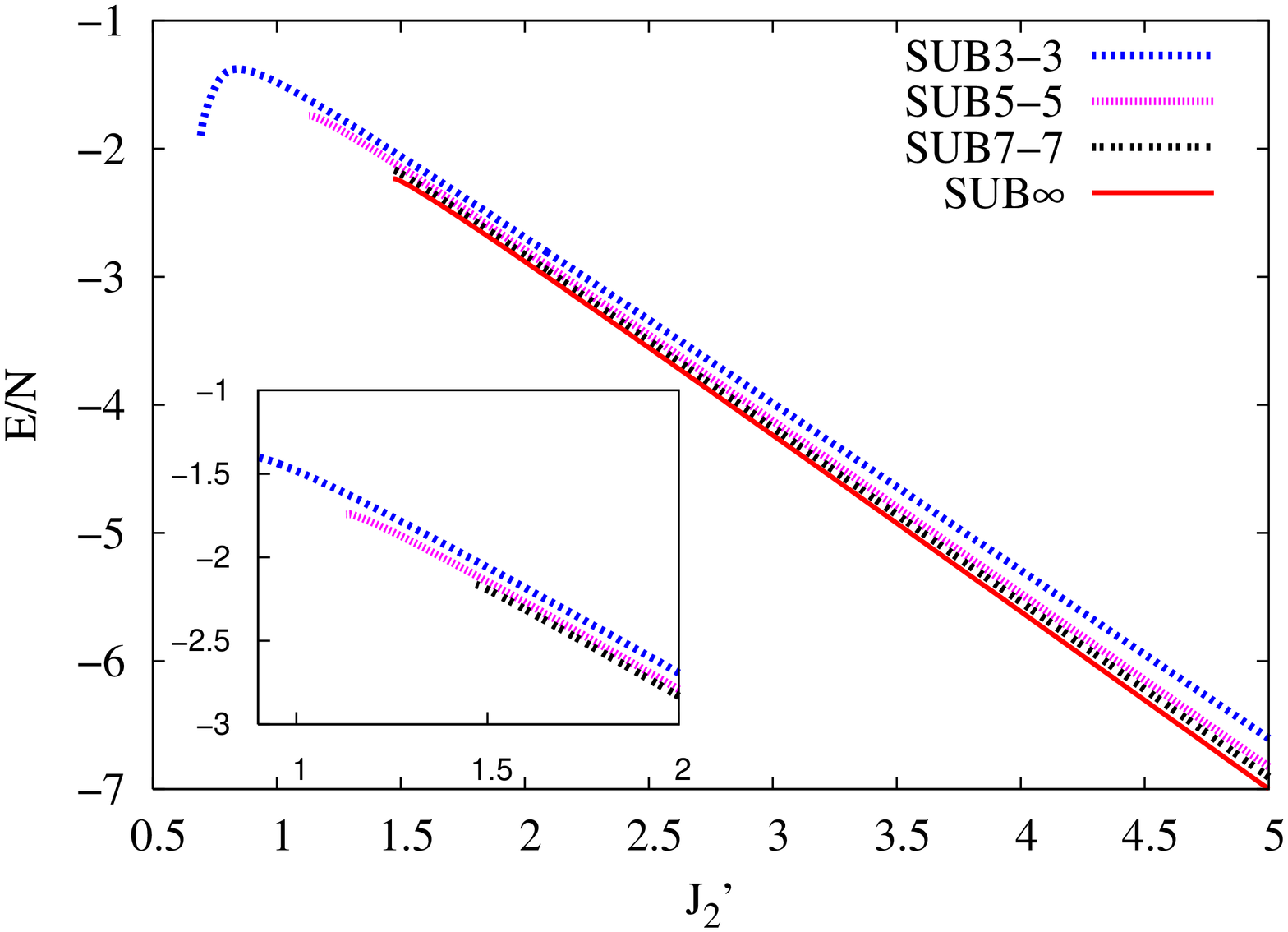}}}
}
\mbox{
\subfloat[$s=\frac{3}{2}$]{\scalebox{0.3}{\includegraphics[angle=270]{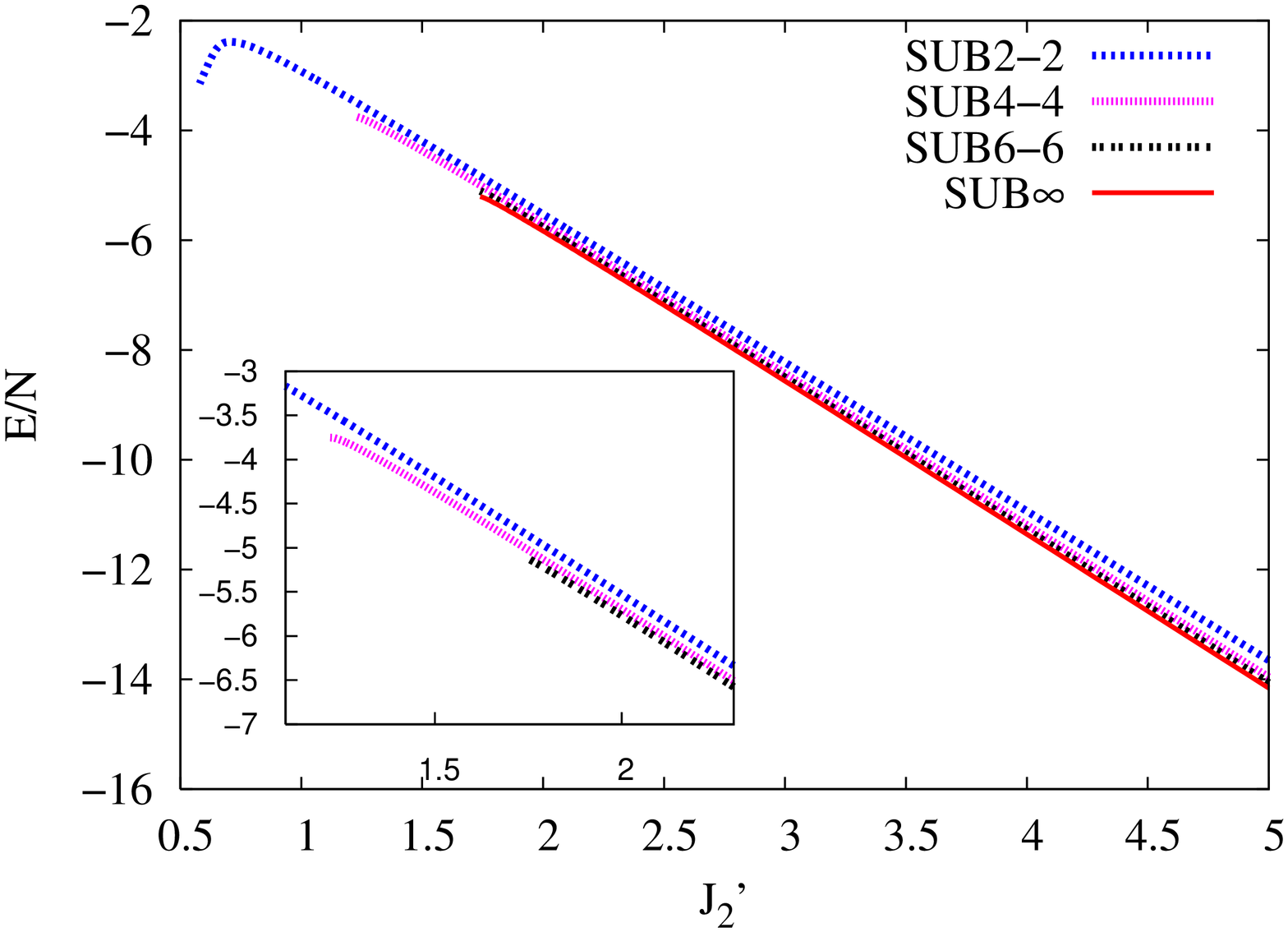}}}
\subfloat[$s=\frac{3}{2}$]{\scalebox{0.3}{\includegraphics[angle=270]{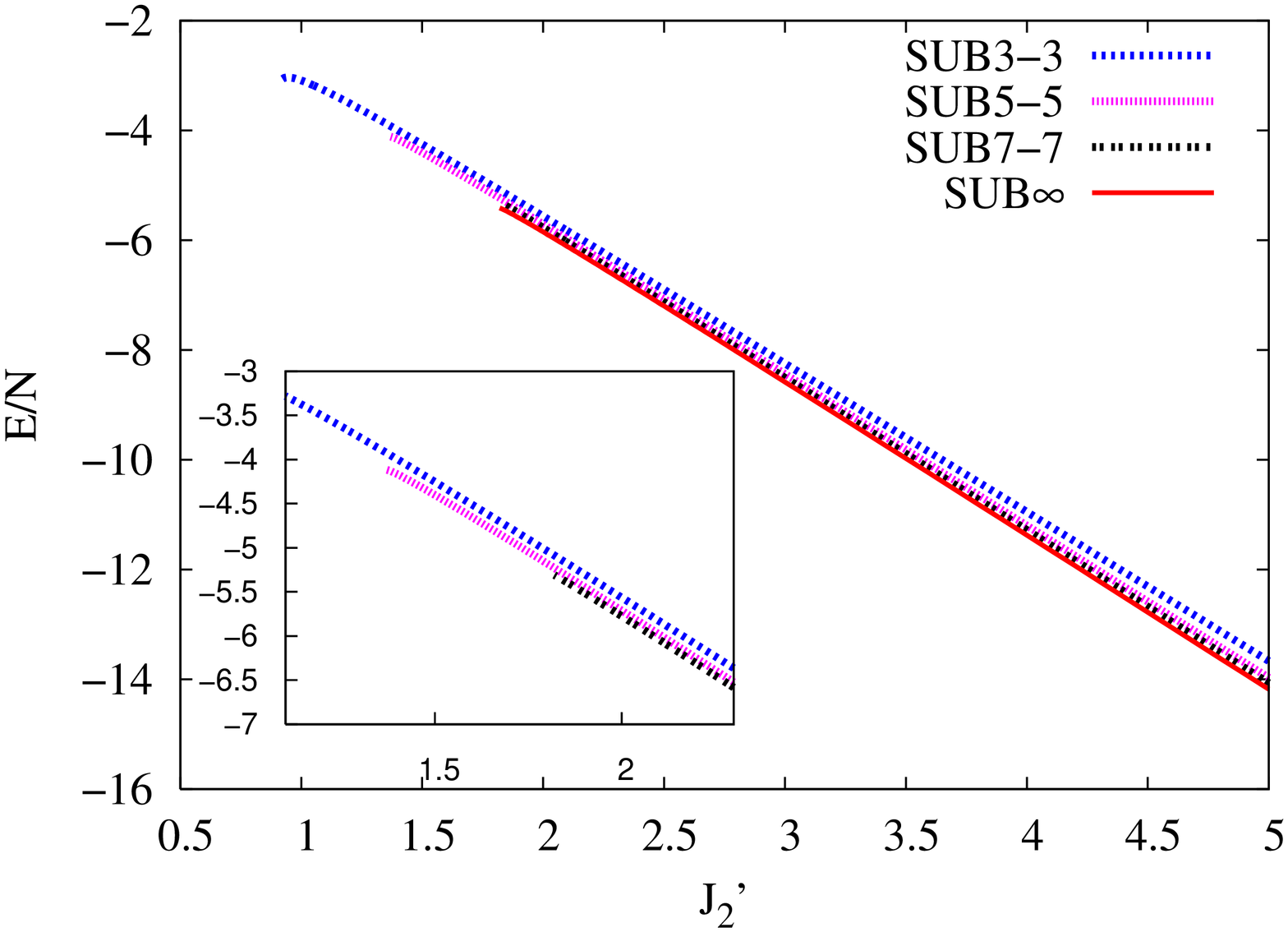}}}
}
\caption{(colour online) Ground-state energy per spin versus $J_{2}'$
  for the stripe-ordered phase of the spin-1 and spin-$\frac{3}{2}$ $J_{1}$-$J_{2}'$ Hamiltonian of Eq.\ (\ref{H})
  with $J_{1}=1$. The CCM results using the
  striped model state are shown for various SUB$n$-$n$ approximations
  ($n=\{2,4,6\}$) and ($n=\{3,5,7\}$). We also show the $n \rightarrow \infty$ extrapolated
  results from using Eq.\ (\ref{Extrapo_E}).}
\label{E_stripe}
\end{center}
\end{figure*}

\begin{figure*}[p!]
\begin{center}
\mbox{
\subfloat[$s=1$]{\scalebox{0.3}{\includegraphics[angle=270]{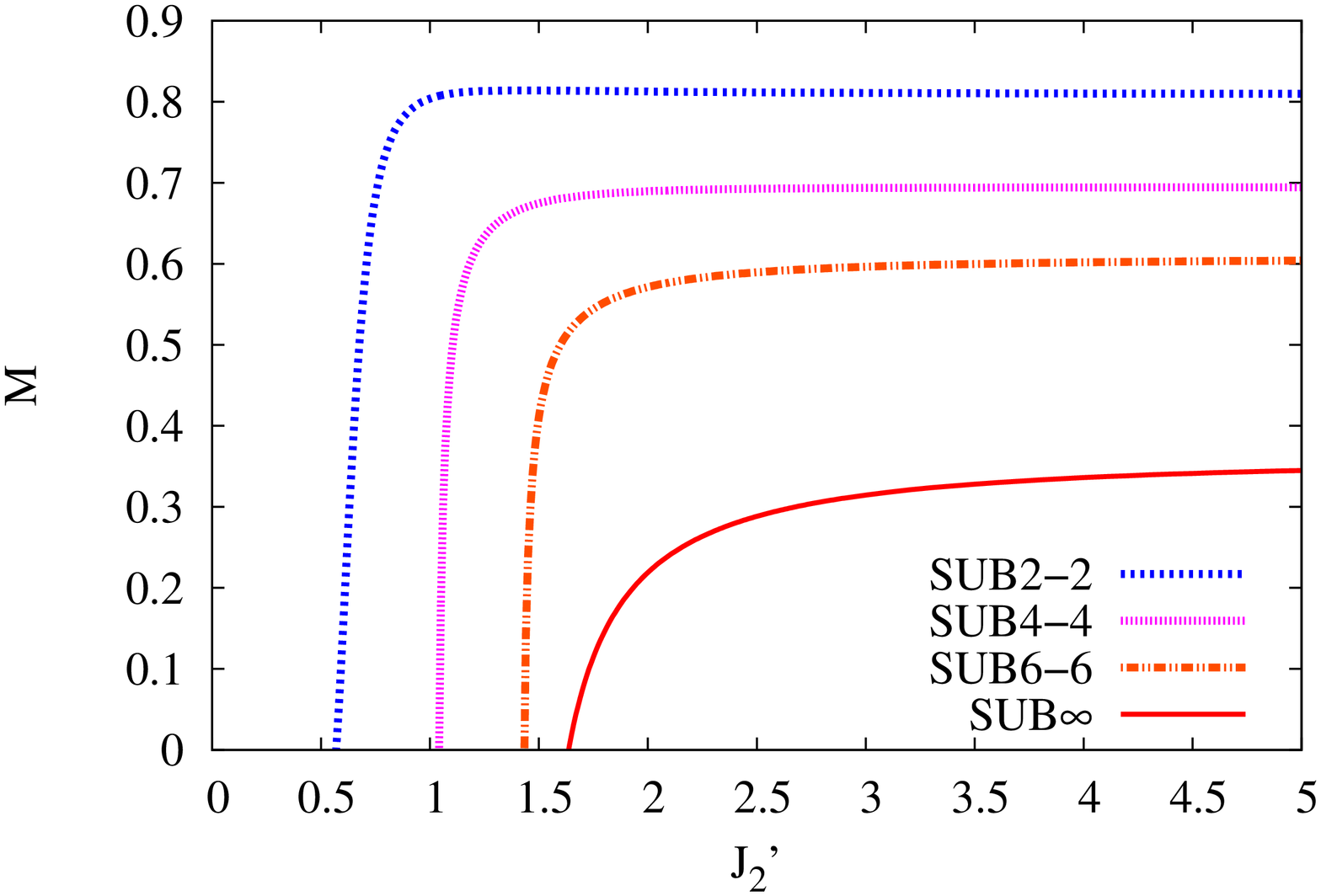}}}
\subfloat[$s=1$]{\scalebox{0.3}{\includegraphics[angle=270]{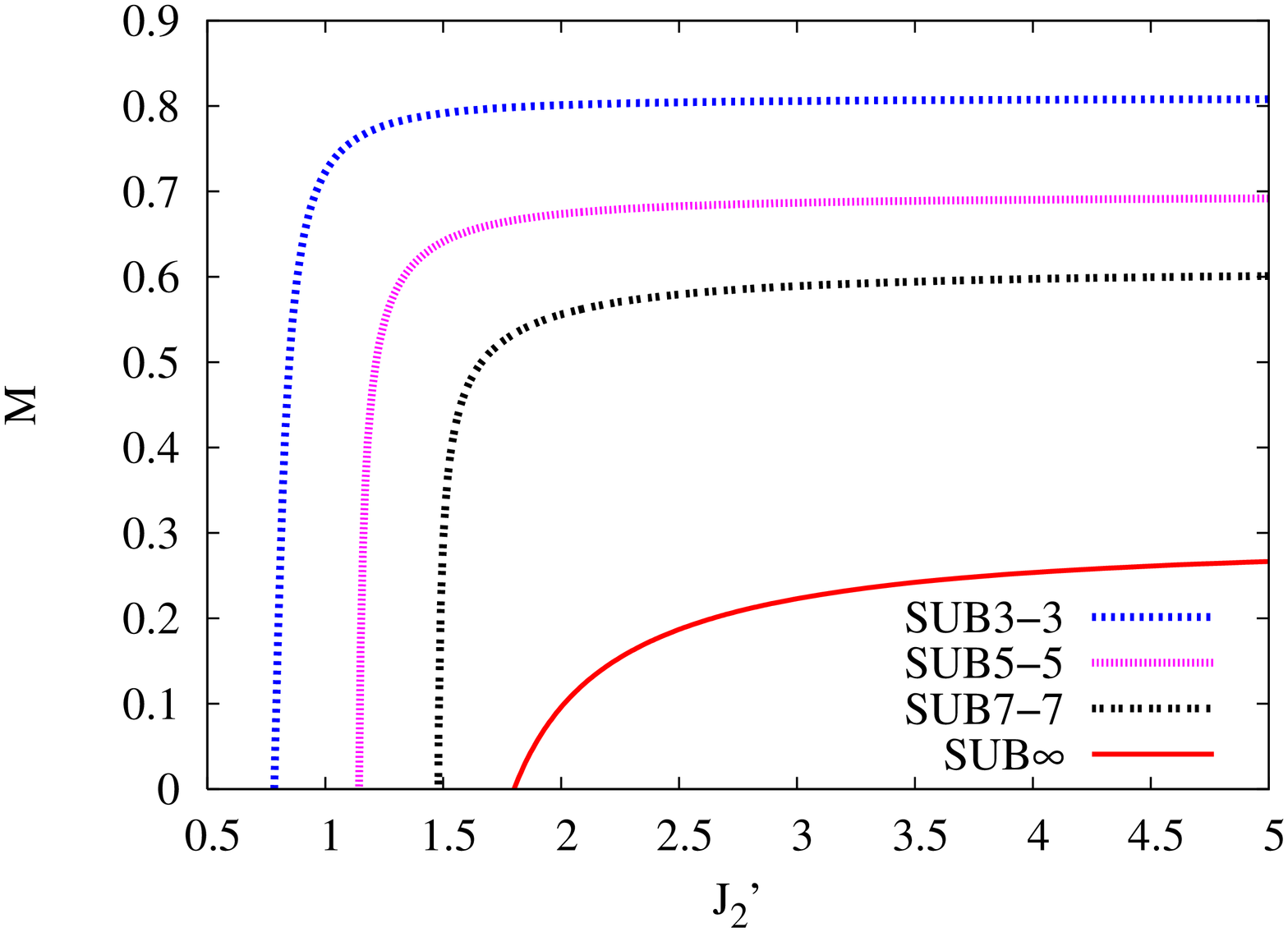}}}
}
\mbox{
\subfloat[$s=\frac{3}{2}$]{\scalebox{0.3}{\includegraphics[angle=270]{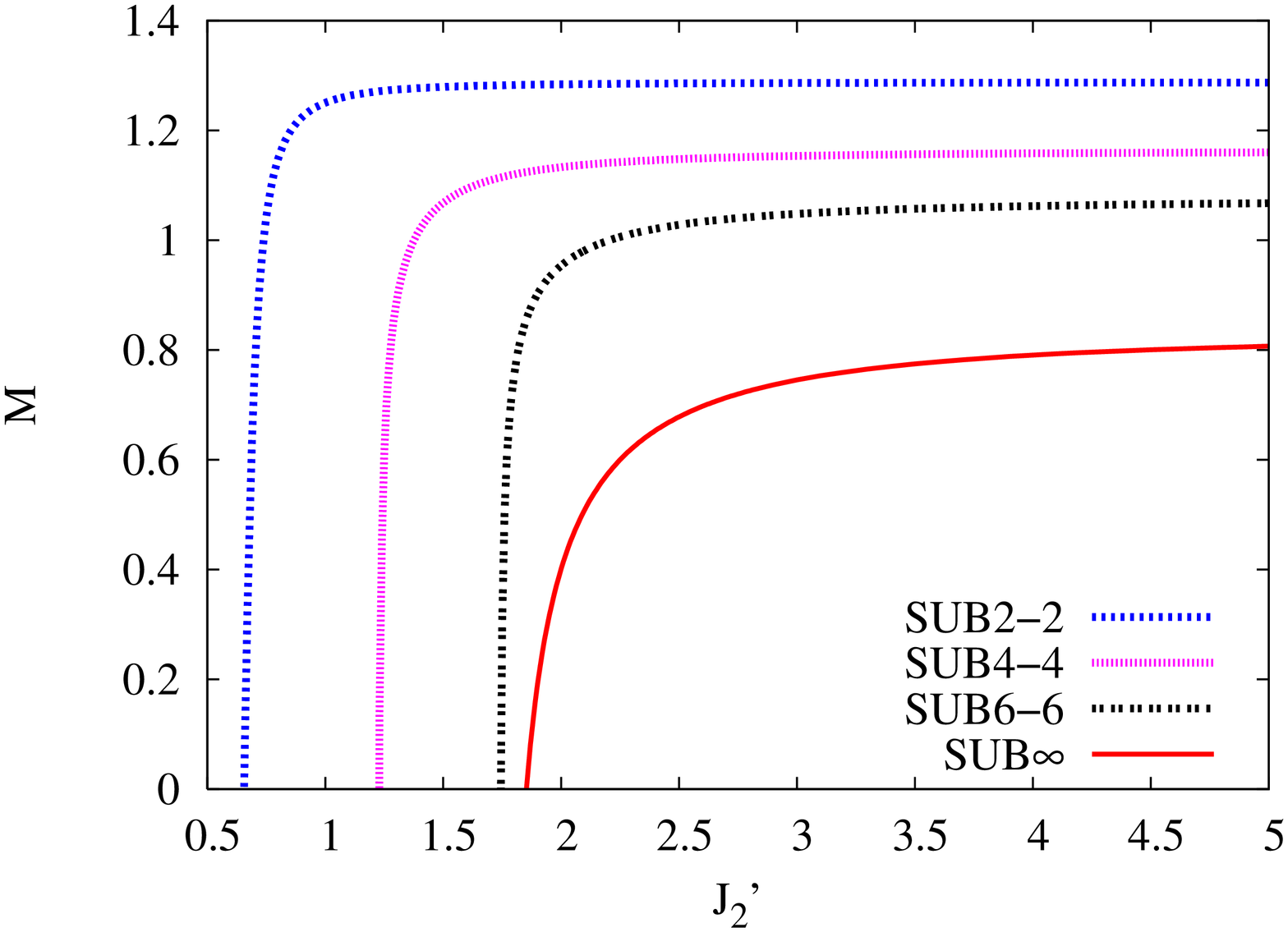}}}
\subfloat[$s=\frac{3}{2}$]{\scalebox{0.3}{\includegraphics[angle=270]{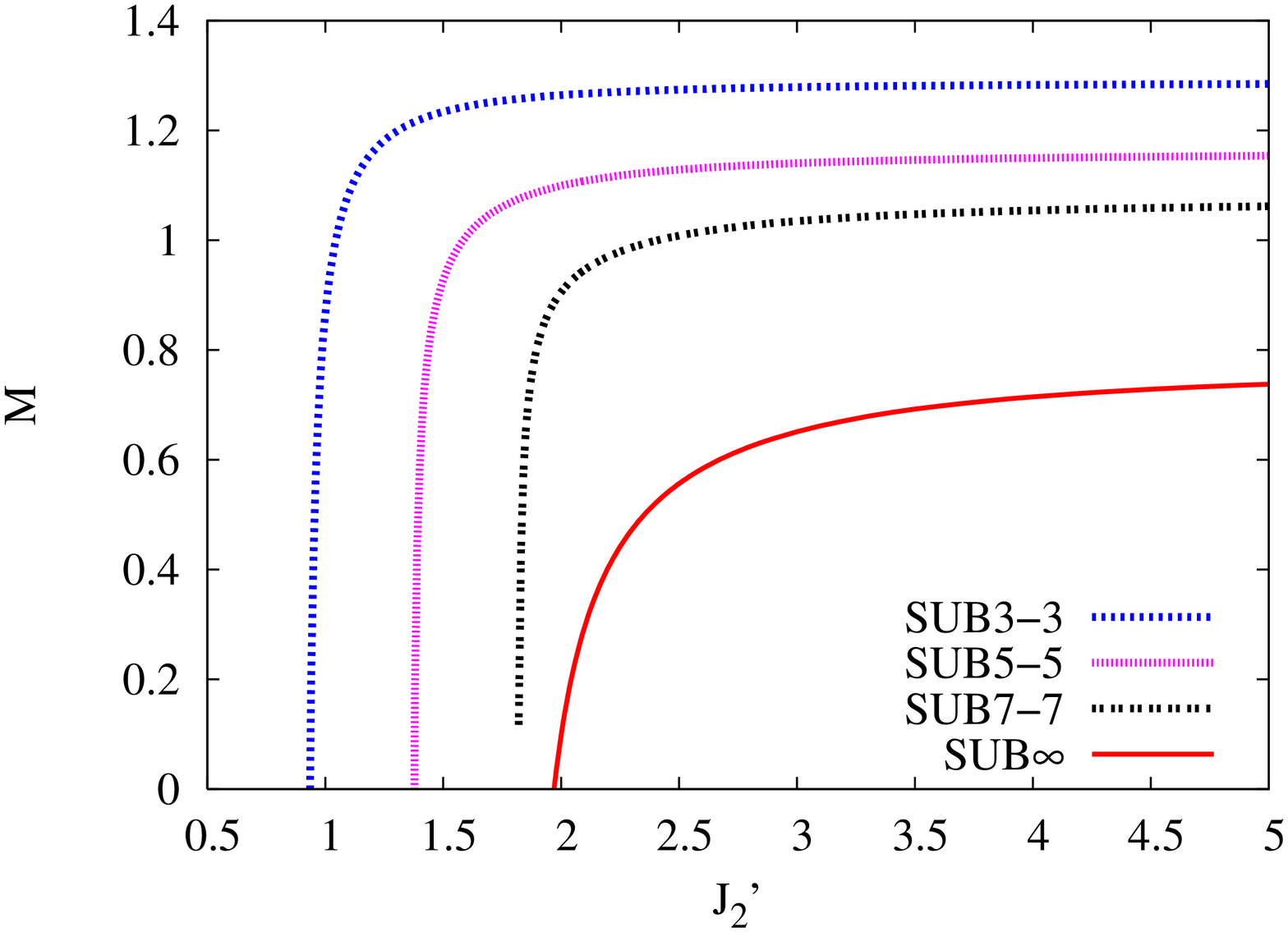}}}
}
\caption{(colour online) Ground-state magnetic order parameter (i.e.,
  the average on-site magnetization) versus $J_{2}'$ for the stripe-ordered
  phase of the spin-1 and spin-$\frac{3}{2}$ $J_{1}$-$J_{2}'$ Hamiltonian of Eq.\
  (\ref{H}) with $J_{1}=1$. The CCM results using the striped model
  state are shown for various SUB$n$-$n$ approximations
  ($n=\{2,4,6\}$ and $n=\{3,5,7\}$). We also show the $n \rightarrow \infty$
  extrapolated results from using Eq.\ (\ref{Extrapo_M}).}
\label{M_stripe}
\end{center}
\end{figure*}

We turn finally to our CCM results based on the collinear striped AFM state 
as the choice for the CCM gs model state $|\Phi \rangle$. 
The SUB$n$-$n$ configurations are again
defined with respect to the triangular lattice geometry, exactly as
before. The numbers of fundamental configuration $N_f$ in each of the 
SUB$n$-$n$ approximations used are given in Table~\ref{FundConf_spin1_SUBnn}. 
Results for the gs energy and magnetic order parameter based
on the striped phase are shown in Figs.\ \ref{E_stripe} and
\ref{M_stripe}
respectively.
%%%%%%%%%%%%%%%%%%%%
%%%%%%%%%%%%%%%%%%%%%%
We see from Fig.\ \ref{E_stripe} that some of the SUB$n$-$n$
solutions based on the striped state for both the $s=1$ and $s=\frac{3}{2}$ cases show a clear
termination point $\kappa_{t}$ of the sort discussed previously, such
that for $\kappa < \kappa_{t}$ no real solution for the striped phase
exists.

\begin{figure*}[tbp!]
\begin{center}
\mbox{
\subfloat[$s=1$]{\scalebox{0.3}{\includegraphics[angle=270]{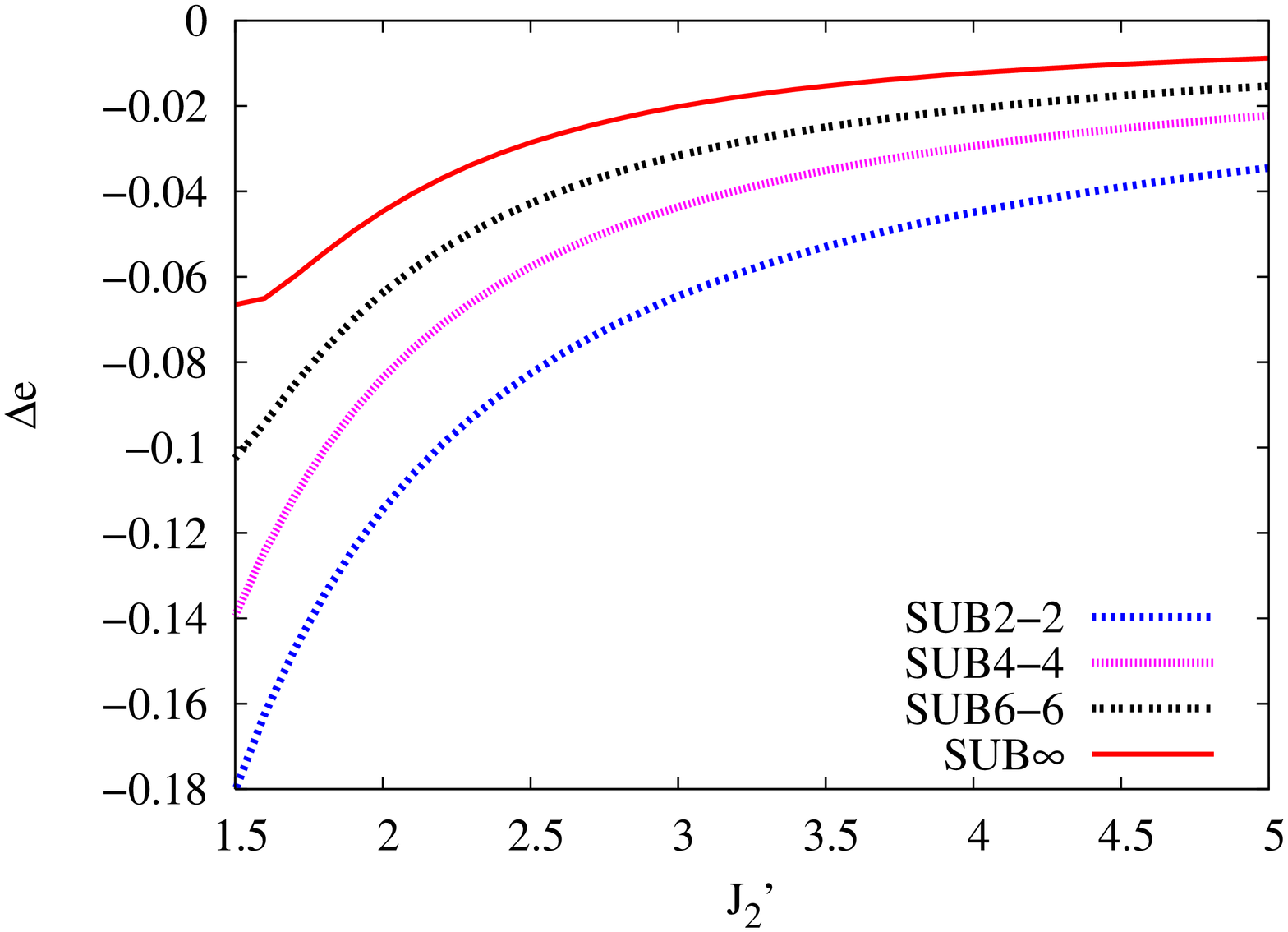}}}
\subfloat[$s=1$]{\scalebox{0.3}{\includegraphics[angle=270]{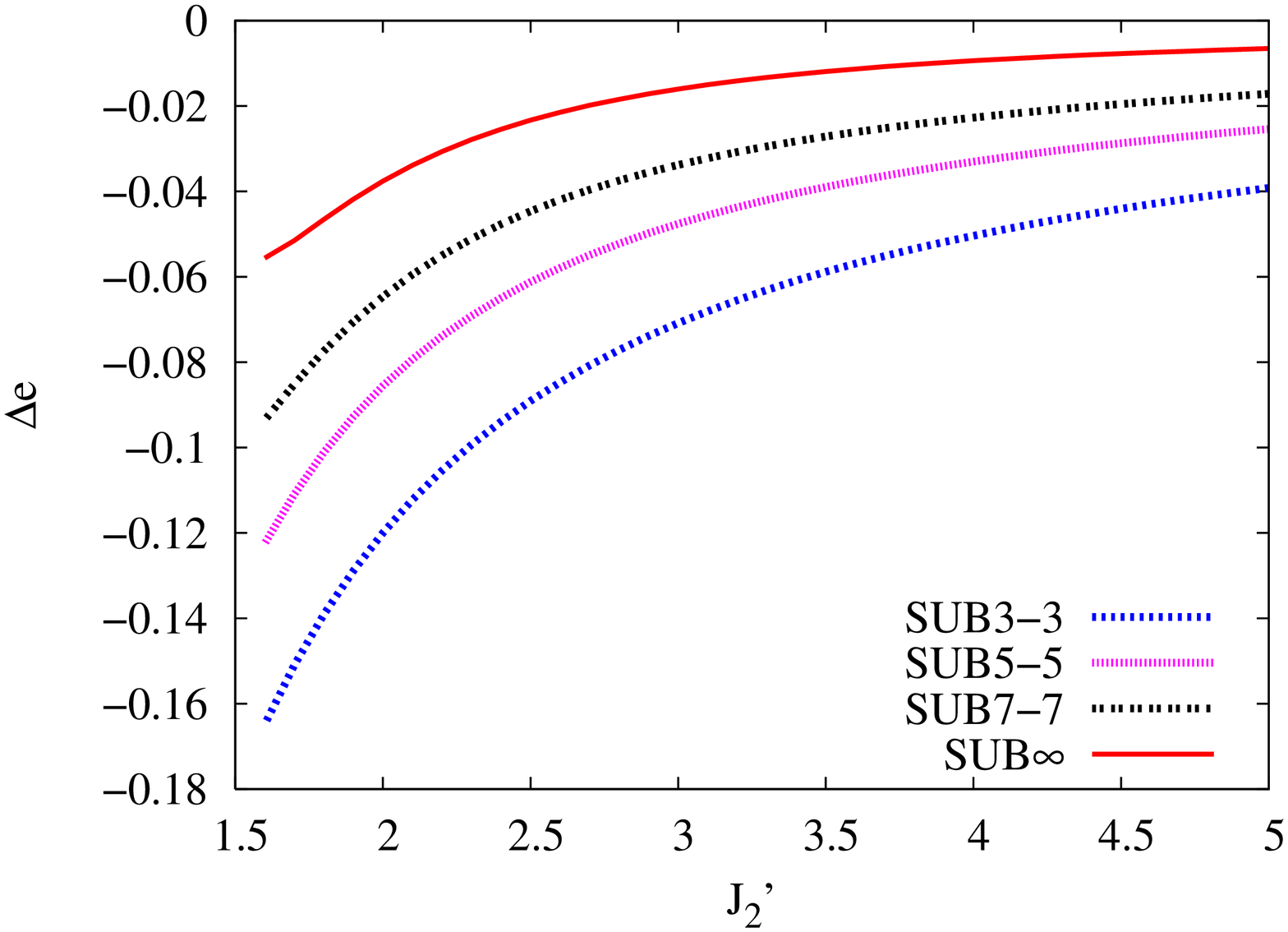}}}
}
\mbox{
\subfloat[$s=\frac{3}{2}$]{\scalebox{0.3}{\includegraphics[angle=270]{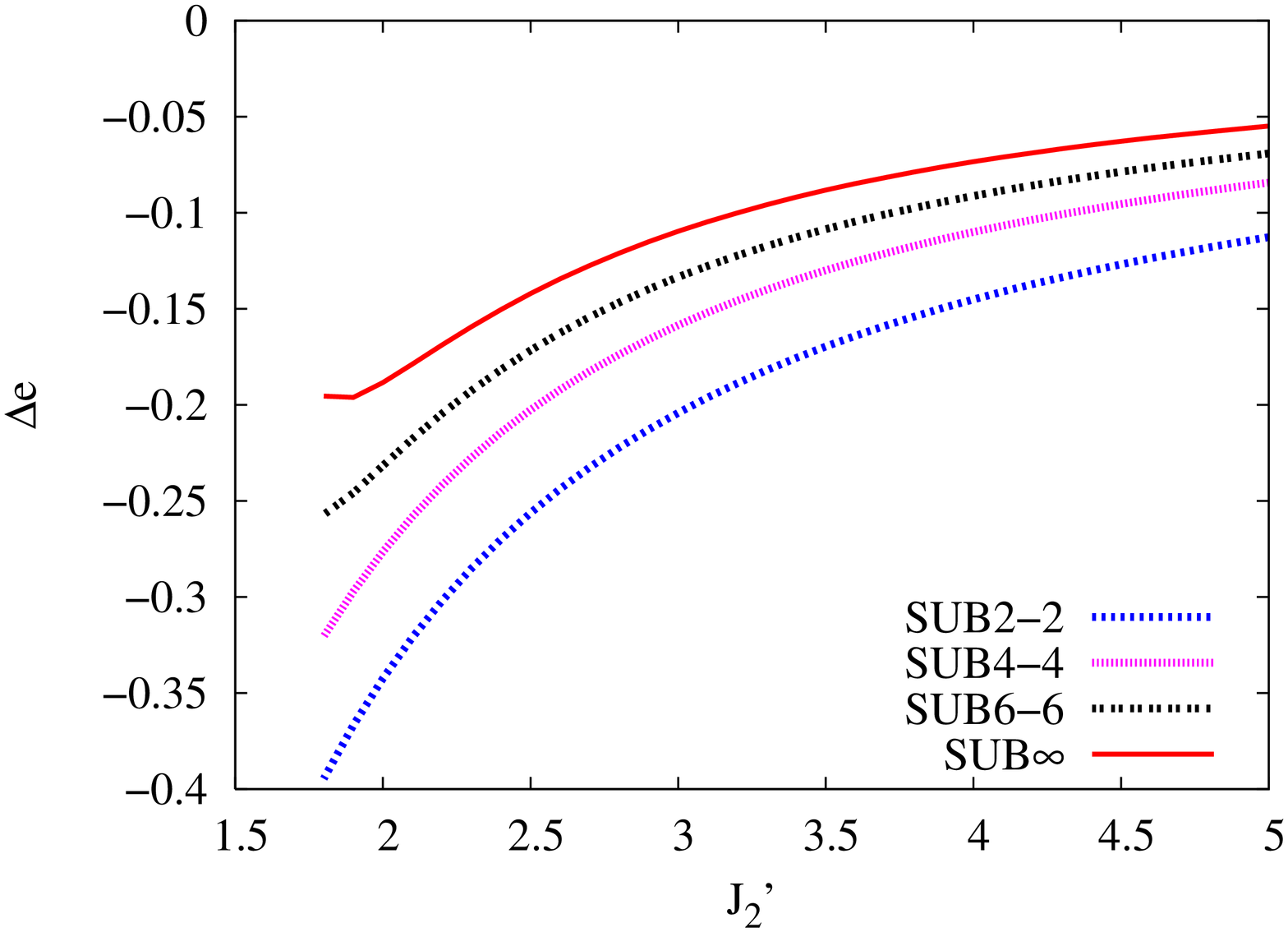}}}
\subfloat[$s=\frac{3}{2}$]{\scalebox{0.3}{\includegraphics[angle=270]{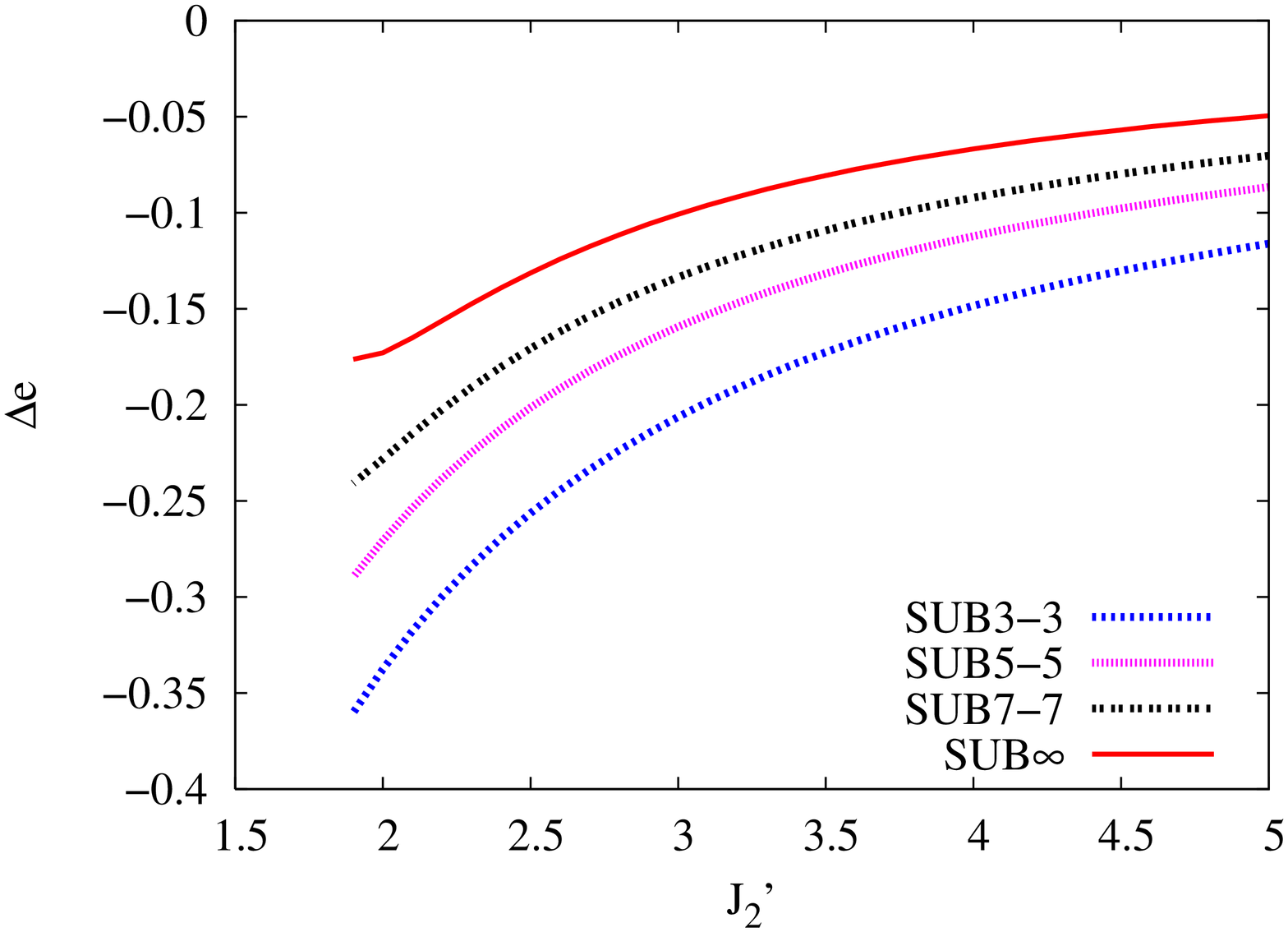}}}
}
\caption{(colour online) Difference between the ground-state energies
  per spin ($e \equiv E/N$) of the spiral and striped phases ($\Delta e
  \equiv e^{{\rm spiral}}- e^{{\rm striped}}$) versus $J_{2}'$ for the
  spin-1 and spin-$\frac{3}{2}$ $J_{1}$-$J_{2}'$ Hamiltonian of Eq.\ (\ref{H})
  with $J_{1}=1$. The CCM results for the energy difference using 
  both the striped and spiral model states for various SUB$n$-$n$
  approximations ($n=\{2,4,6\}$) and ($n=\{3,5,7\}$) are shown. We
  also show the $n \rightarrow \infty$ extrapolated results from using
  Eq.\ (\ref{Extrapo_E}) for the two phases separately.}
\label{EDiff}
\end{center}
\end{figure*}

For the spin-$1$ model the large-$\kappa$ limit of the extrapolated
SUB$n$-$n$ energy per spin results of
$E/N=-1.3897J_{2}'$ from Fig.\ \ref{E_stripe}(a) using $n=\{2,4,6\}$ and $E/N=-1.3936J_{2}'$ 
from Fig.\ \ref{E_stripe}(b) using $n=\{3,5,7\}$ agree well
with the known 1D chain result of $E/N=-1.4015$ obtained from a
density-matrix renormalization group analysis \cite{Wh:1993} and 
our previous CCM result \cite{Fa:2002}, just as
in Fig.\ \ref{E}(a) and (b) for the spiral phase. Similarly, for the
spin-$\frac{3}{2}$ case, the large-$\kappa$ limit of the 
extrapolated SUB$n$-$n$ results for the energy per
spin of $E/N=-2.8205J_{2}'$ from Fig.\ \ref{E_stripe}(c) using $n=\{2,4,6\}$ and
$E/N=-2.8243J_{2}'$ from Fig.\ \ref{E_stripe}(d) using $n=\{3,5,7\}$, 
with almost identical results again obtained from Fig.\ \ref{E}(c) and (d). 
Unlike for their spin-$\frac{1}{2}$ counterpart, however, 
the striped phase is never a stable gs state
for either the spin-1 or spin-$\frac{3}{2}$ models, because their
energies always lie higher than those of the spiral state for all values of
$J_{2}'$, as shown in Fig.~\ref{EDiff}.
Hence for the $s=1$ and $s=\frac{3}{2}$ cases,
there is only one quantum critical point $\kappa_{c}$, at which 
the N\'{e}el phase is driven to the helical phase.

\section{Discussion and conclusions}
\label{discussion}
In an earlier paper~\cite{Bi:2008_SqTrian} we used the CCM to study
the effect of quantum fluctuations on the zero-temperature gs phase
diagram of a frustrated spin-$\frac{1}{2}$ interpolating
square-triangle antiferromagnetic model.  This is the so-called
$J_{1}$--$J_{2}'$ model, defined on an anisotropic 2D lattice, as 
shown in  Fig.~\ref{model}.  In the current paper we have extended 
the analysis to consider spin-1 and spin-$\frac{3}{2}$ versions of the 
same model.  As before we have studied the case where the NN
$J_{1}$ bonds are antiferromagnetic ($J_{1} > 0$) and the competing
$J_{2}' \equiv \kappa J_{1}$ bonds have a strength $\kappa$ that
varies from $\kappa = 0$ (corresponding to the HAF on the square
lattice) to $\kappa \rightarrow \infty$ (corresponding to a set of
decoupled 1D HAF chains), with the HAF on the triangular lattice as
another special case, $\kappa = 1$, in between the two extremes. The results
of the $\kappa=0$ limit of the present model (and see
Table~\ref{EandM_spiral}) for the $s=1$ case are comparable with
those obtained from the SWT and SE techniques
\cite{Ha:1992,Zh:1991} which are among the best
alternative numerical method to the CCM for highly frustrated
spin-lattice models like the present $J_{1}$--$J_{2}'$ model.

For the spin-1 model we find that the phase transition between the
N\'{e}el antiferromagnetic phase and the spiral phase
occurs at the value $\kappa_{c} = 0.615 \pm 0.010$, whereas for the
spin-$\frac{3}{2}$ model we find that the phase transition occurs at
$\kappa_{c} = 0.575 \pm 0.005$.  From the continuous and smooth behaviour 
of the energies of the two phases it appears that the transition is
second-order, as in the classical case. However, 
on neither side of the transition at $\kappa_{c}$
does the order parameter $M$ (i.e., the average on-site magnetization)
go to zero for either of the two higher spins considered here.  On
the other hand, unlike in the spin-$\frac{1}{2}$ case, in neither of the
higher-spin models does there appear to be any discontinuity in $M$ at the 
transition.  All of the indications are thus that the transition
between the N\'{e}el antiferromagnetic and the spiral phases is of continuous 
(second-order) type for both cases $s=1$ and $s=\frac{3}{2}$, in 
contrast to the spin-$\frac{1}{2}$ case where the order parameter $M$ 
appeared to show a discontinuous jump at the transition, which was
found to be a weakly first-oder one (although it could not be entirely 
excluded on the available evidence that the transition might be 
a second-order one).

We have observed that as the quantum
spin number $s$ is increased, the position of the quantum critical point
at $\kappa_{c}$ between the phases with N\'{e}el and spiral order 
is brought closer to the classical ($s \rightarrow
\infty$) value, $\kappa_{{\rm cl}}=0.5$, as expected.  In contrast with the
$s=\frac{1}{2}$ case where there is a second quantum critical point
for the phase transition from the helical phase to a collinear stripe-ordered
phase, we find no evidence at all for such a further transition for
either of the cases $s=1$ or $s=\frac{3}{2}$.

We note that the spin-1 HAF on the (undistorted) triangular lattice (viz., our 
limiting case $\kappa = 1$) has itself been the subject of much recent interest 
from both the theoretical and experimental viewpoints.  From the experimental side 
spin-1 models on the triangular lattice are believed to underlie the properties 
of such materials as NiGa$_{2}$S$_{4}$ \cite{Nakatsuji:2005} and 
Ba$_{3}$NiSb$_{2}$O$_{9}$ \cite{Cheng:2011}.  In both materials the Ni$^{2+}$ 
ions form in weakly coupled 2D triangular lattice layers.
Thus, for example, thermodynamic and neutron scattering measurements 
on NiGa$_{2}$S$_{4}$ show conclusive evidence that the inherent geometric 
frustration of the triangular lattice stabilises a low-temperature 
spin-disordered state, which was proposed as being consistent with a spin-liquid 
phase \cite{Nakatsuji:2005}.  Other candidates for spin-1 quantum spin-liquid 
phases have more recently been proposed from an experimental study of the 
high-pressure sequence of structural phases in the material 
Ba$_{3}$NiSb$_{2}$O$_{9}$ \cite{Cheng:2011}.  

Whereas quantum fluctuations are certainly intrinsically greatest for 
spin-lattice systems with the lowest spin value $s=\frac{1}{2}$, as we have also
found here, such effects can also be enhanced for the $s>\frac{1}{2}$ cases by 
the addition to the pure (bilinear) Heisenberg interaction with NN terms only 
of terms such as a NN biquadratic interaction or other higher-order exchange 
terms.  It is precisely by the addition of terms like this that unusual quantum 
ground states, such as ones with quadrupolar (or spin-nematic) order have been 
predicted theoretically to be stabilised for the spin-1 HAF on the triangular
lattice \cite{Bhatta:2006,Lauchli:2006,Tsunetsugu:2006,Stoudenmire:2009}.  It 
is argued that such a state can account for the observed low-temperature 
thermodynamics in the spin-1, quasi-2D, antiferromagnetic material NiGa$_{2}$S$_{4}$,
although at the lowest temperatures the observed order is that of an
(incommensurate) spiral phase.  In a very recent paper (that appeared only after
submission of this paper) \cite{Xu:2011} it is also argued that the quantum
spin-liquid phases presumed to have been seen in recent experiments 
\cite{Cheng:2011} in the layered material Ba$_{3}$NiSb$_{2}$O$_{9}$, may 
be explained microscopically as emanating from a spin-1 HAF on the triangular 
lattice with both NN and NNN isotropic antiferromagnetic Heisenberg couplings.

Such other (e.g., spin-1) models on the triangular lattice as those described above,
involving either isotropic bilinear and biquadratic couplings or both NN and NNN
bilinear Heisenberg couplings, could also be investigated via the CCM, and it 
would surely be interesting to do so.  While such additional terms in the
Hamiltonian present no additional obstacles to the use of the method at all, the 
choice of which model states to use always enters at the outset. It is certainly 
true that most calculations on spin systems employing the CCM, including those in 
the present paper, employ model states built by independent-spin product states 
for which the choice of state for the spin on each site is formally independent of 
the choice of all others.  Often for these independent-spin product model states the 
use of collinear states, such as the N\'{e}el or striped states considered here, 
is possible, where all spins are aligned parallel or antiparallel to one axis.  However, 
as we have seen, noncollinear (e.g. spiral) model states can sometimes be favourable 
for certain values of the frustration.  In either case multispin correlations are 
then included systematically on top of the independent-spin product
model states.  As we have seen here, the CCM for such independent-spin 
product model states may then be applied to high orders by
using a computational implementation used here and described more fully elsewhere 
(see, e.g., Refs.~\cite{Ze:1998,Fa:2002,ccm} and references cited therein).
In particular, it may be applied to lattices of complex crystallographic symmetry. 
Furthermore, as seen here, it is not constrained to systems with spin quantum 
number $s = \frac{1}{2}$.

When the system under consideration may have more exotic ground states with less
conventional ordering than the (often essentially quasiclassical) independent-spin 
product states described above, the CCM may still be very profitably employed.
Even the use of such independent-spin product states can still give very precise
phase boundaries for when such states give way to more exotic states.  A good 
example among many to date is the well-studied frustrated spin-$\frac{1}{2}$
$J_{1}$-$J_{2}$ model on the square lattice discussed in Sec.~\ref{Intro}, for 
which the phase boundaries of the non-classical paramagnetic state 
(that has no magnetic LRO) have been estimated very  
accurately using the CCM in Refs. \cite{Bi:2008_JPCM_V20_p255251,Bi:2008_PRB}.  
A more recent example is provided by a CCM calculation \cite{FBLRC:2011} 
of the frustrated spin-$\frac{1}{2}$
$J_{1}$-$J_{2}$-$J_{3}$ model on the honeycomb lattice which incorporates 
NN bonds ($J_{1}$) and NNN bonds ($J_{2}$) as in the $J_{1}$-$J_{2}$ model, but
now also includes next-next-nearest-neighbour bonds ($J_{3}$).  For the case
$J_{3}=J_{2}$ an intermediate paramagnetic phase was accurately located between
collinear antiferroimagnetic states of quasiclassical N\'{e}el and striped order.  By 
calculating with such model states the plaquette susceptibility, the authors gave 
precise values not only of the phase boundaries of this intermediate state, but also
gave clear evidence that it had plaquette valence-bond crystalline ordering.

It is also worth noting that the CCM can deal directly with more complex model 
states, such as those involving valence-bond crystal (VBC) order.  Thus, for example,
non-classical VBC ordering has been considered using the CCM
by employing directly valence-bond model states, i.e. two- or multi-spin singlet product
states \cite{Xian:1994}.  A drawback of this approach is that it involves the 
direct use of products of localized states (e.g., two-spin dimers or multi-spin 
plaquettes) in the model state.  Hence, this approach requires that a new 
matrix-operator formalism be created for each new problem.  Also,
the Hamiltonian and CCM ket- and bra-state operators must be written in terms 
of this new matrix algebra.  The CCM equations may be derived and solved 
once the commutation relationships between the operators have been established. 
Although formally straightforward, this process can be tedious and time-consuming. 
Furthermore, the existing high-order CCM formalism and codes also need to be 
amended extensively for each separate model considered.

More recently a quite different CCM approach has been advocated for dealing with
such VBC states \cite{Farnell:2009}.  It starts directly from
collinear independent-spin product model states, and shows how one may form 
exact local dimer or plaquette ground states within the CCM framework.  This approach 
has the huge advantages of being conceptually simple and thus also of being easy 
to implement.  Furthermore, one may then use directly the existing high-order CCM 
formalism, computer codes, and extrapolation schemes used and described here and 
in that references cited.  To date the method has been applied with excellent 
results to the spin-$\frac{1}{2}$ $J_{1}$-$J_{2}$ model for the linear chain, 
the spin-$\frac{1}{2}$ Shastry-Sutherland model on the 2D square lattice \cite{SS:1981}, 
and the so-called spin-$\frac{1}{2}$ $J$-$J'$ HAF on the 2D CAVO lattice that
is appropriate to the magnetic material CaV$_{4}$O$_{9}$.  It is a one-fifth 
depleted square lattice, and the model on this lattice comprises 
two nonequivalent antiferromagnetic NN bonds of strength $J$ and $J'$.  The $J$ bonds
connect sites on the NN four-spin square plaquettes while the $J'$ (dimer) bonds connect
NN sites belonging to neighbouring square plaquettes.

In conclusion, it will be of interest to use the CCM for the other spin-1 models 
discussed above on the triangular lattice that are believed to be relevant to such
quasi-2D materials as NiGa$_{2}$S$_{4}$ and Ba$_{3}$NiSb$_{2}$O$_{9}$, for both 
of which considerable experimental data exist.  We hope to be able to perform and
report ourselves on such calculations at a later date.

\section*{Acknowledgment}
We thank the University of Minnesota Supercomputing Institute for
Digital Simulation and Advanced Computation for the grant of
supercomputing facilities, on which we relied heavily for the
numerical calculations reported here.  We also thank D.~J.~J. Farnell
and C.~E. Campbell for their assistance.

%
% BibTeX users please use
% \bibliographystyle{}
% \bibliography{}

\begin{thebibliography}{99}

\bibitem{Sa:1995}
S.~Sachdev, in {\it Low Dimensional Quantum Field Theories for Condensed Matter Physicists},
edited by~Y.~Lu, S.~Lundqvist, and G. Morandi (World Scientific, Singapore 1995).

\bibitem{Ri:2004} 
J.~Richter, J.~Schulenburg, and A.~Honecker, 
in {\em Quantum Magnetism}, Lecture Notes in
Physics {\bf 645}, edited by~U.~Schollw{\"{o}}ck, J.~Richter,
D.~J.~J.~Farnell, and R.~F.~Bishop (Springer-Verlag, Berlin, 2004), p.~85.

\bibitem{Mi:2005} 
G.~Misguich and C.~Lhuillier, 
in {\em Frustrated Spin Systems}, edited by H.~T.~Diep (World Scientific, Singapore, 2005), p.~229.

\bibitem{ccm_UJack_asUJ_2010}
R.~F.~Bishop, P.~H.~Y.~Li, D.~J.~J.~Farnell, and C~E.~Campbell,
Phys.\ Rev.\ B {\bf 82}, 024416 (2010).

\bibitem{Da:2005_JPhy_17} 
R.~Darradi, J.~Richter, and D.~J.~J~Farnell,
J. Phys.: Condens. Matter, {\bf 17}, 341 (2005).

\bibitem{Ha:1983} 
F.~D.~M.~Haldane, 
Phys.\ Lett.\ A {\bf 93} 464 (1983);
Phys.\ Rev.\ Lett.\ {\bf 50}, 1153 (1983).  

\bibitem{Vo:2001}
A.~Voigt, J.~Richter, and P.~Tomczak,
Physica A {\bf 299}, 461 (2001).

\bibitem{Bi:1992}
R.~F.~Bishop, J.~B.~Parkinson, and Y.~Xian,
Phys.\ Rev.\ B {\bf 46}, 880 (1992).

\bibitem{Fa:2002} 
D.~J. J.~Farnell, R.~F.~Bishop, and K.~A.~Gernoth, 
J.\ Stat.\ Phys.\ {\bf 108}, 401 (2002).

\bibitem{Grover:2010_s1}
T.~Grover and T.~Senthil,
arXiv:1012.5669v1 [cond-mat.str-el] (2010).

\bibitem{Lin:1989}
H.~Q.~Lin and V.~J.~Emery,
Phys.\ Rev.\ B {\bf 40}, 2730 (1989).

\bibitem{Ir:1992}
V.~Y.~Irkhin, A.~A.~Katanin, and M.~I.~Katsnelson,
J.\ Phys.: Condens.\ Matter {\bf 4}, 5227 (1992).

\bibitem{Fa:2001_PRB64} 
D.~J.~J.~Farnell, K.~A.~Gernoth, and R.~F.~Bishop, 
Phys.\ Rev.\ B {\bf 64}, 172409 (2001).

\bibitem{Mo:2006} 
S.~Moukouri,
J.\ Stat.\ Mech.\ P02002, (2006).
      
\bibitem{Bi:2008_EPL} 
R.~F.~Bishop, P.~H.~Y.~Li, R.~Darradi, and J.~Richter, 
Europhys.\ Lett.\ {\bf 83}, 47004 (2008) 

\bibitem{Bi:2008_JPCM_V20_p415213} 
R.~F.~Bishop, P.~H.~Y.~Li, R.~Darradi, J.~Richter, and C.~E.~Campbell, 
J.\ Phys.: Condens.\ Matter {\bf 20}, 415213 (2008).

\bibitem{Ji:2009} 
H.~C.~Jiang, F.~Kr\"{u}ger, J.~E.~Moore, 
D.~N.~Sheng, J.~Zaanen, and Z.~Y.~Weng, 
Phys.\ Rev.\ B {\bf 79}, 174409 (2009).

\bibitem{Bishop:2010_UJack_GrtSpins}
R.~F.~Bishop and P.~H.~Y.~Li,
Eur. Phys. J. B {\bf 81}, 37 (2011).
%arXiv:1010.5161v1 [cond-mat.str-el] (2010).

\bibitem{Zhao:2011_honeycomb_s1}
H.~H.~Zhao, Q.~N.~Chen, Z.~C.~Wei, M.~P.~Qin, G.~M.~Zhang, and T. Xiang,
arXiv:1105.2716v1 [cond-mat.str-el] (2011).

\bibitem{Vr:2008} 
M.~A.~de Vries, T.~K.~Johal, A.~Mirone, 
J.~S.~Claydon, G.~J.~Nilsen, H.~M.~R{\o}nnow, G.~van der Laan, and
A.~Harrison, 
Phys.\ Rev.\ B {\bf 79}, 045102 (2009).

\bibitem{KWHH:2008} 
Y.~Kamihara, T.~Watanabe, M.~Hirano, and H.~Hosono,
J.\ Am.\ Chem.\ Soc.\ {\bf 130}, 3296 (2008).

\bibitem{Ma:2008} 
F.~Ma, Z.-Y.~Lu, and T.~Xiang,
Phys.\ Rev.\ B {\bf 78}, 224517 (2008).       
   
\bibitem{Bi:1991} 
R.~F.~Bishop, 
Theor.\ Chim.\ Acta {\bf 80}, 95 (1991).

\bibitem{Bi:1998} 
R.~F.~Bishop,  
in {\em Microscopic Quantum Many-Body Theories and Their Applications}, 
edited by J.~Navarro and A.~Polls, {\em Lecture Notes in Physics} {\bf 510} 
(Springer-Verlag, Berlin, 1998), p.1.

\bibitem{Fa:2004} 
D.~J.~J.~Farnell and R.~F.~Bishop, 
in {\em Quantum Magnetism}, 
edited by U.~Schollw{\"{o}}ck, J.~Richter, D.~J.~J.~Farnell, and R.~F.~Bishop, 
{\em Lecture Notes in Physics} {\bf 645} (Springer-Verlag, Berlin, 2004), p.307.

\bibitem{Bi:2008_SqTrian} 
R.~F.~Bishop, P.~H.~Y.~Li, D.~J.~J.~Farnell, and C.~E.~Campbell, 
Phys.\ Rev.\ B {\bf 79}, 174405 (2009).

\bibitem{Bi:2010_SqTrian_IJMPB}
R.~F.~Bishop, P.~H.~Y.~Li, D.~J.~J.~Farnell, and C.~E.~Campbell,
Int.\ J.\ Mod.\ Phys. B {\bf 24}, 5011 (2010); {\it ibid}. Erratum (2011).

\bibitem{Me:1999}
J.~Merino, R.~H.~McKenzie, J.~B.~Marston, and C.~H.~Chung,
J.\ Phys.: Condens.\ Matter {\bf 11}, 2965 (1999).

\bibitem{We:1999}
Zheng Weihong, R.~H.~McKenzie, and R.~R.~P.~Singh,
Phys.\ Rev.\ B {\bf 59}, 14367 (1999).

\bibitem{St:2007}
O.~A.~Starykh and L.~Balents,
Phys.\ Rev.\ Lett.\ {\bf 98}, 077205 (2007).

\bibitem{Pa:2008}
T.~Pardini and R.~R.~P.~Singh,
Phys.\ Rev.\ B {\bf 77}, 214433 (2008).

\bibitem{Ki:1996}
H.~Kino and H.~Fukuyama,
J.\ Phys.\ Soc. Japan {\bf 65}, 2158 (1996);
R.~H.~McKenzie, 
Comments Condens.\ Matter Phys.\ {\bf 18}, 309 (1998).

\bibitem{Co:1997} 
R.~Coldea, D. A.~Tennant, R.~A.~Cowley, D.~F.~McMorrow, B.~Dorner, and Z.~Tylczynski, 
Phys.\ Rev.\ Lett. {\bf 79}, 151 (1997).

\bibitem{Be:1931}
H.~A.~Bethe, 
Z.\ Phys.\ {\bf 71}, 205 (1931).

\bibitem{Vi:1977} 
J.~Villain, 
J.\ Phys.\ (France) {\bf 38}, 385 (1977); 
J.~Villain, R.~Bidaux, J.~P.~Carton, and R.~Conte, 
{\it ibid.} {\bf 41}, 1263 (1980);
E.~Shender, 
Sov.\ Phys.\ JETP {\bf 56}, 178 (1982).

\bibitem{Bi:2008_JPCM_V20_p255251} 
R.~F.~Bishop, P.~H.~Y.~Li, R.~Darradi, and J.~Richter, 
J.\ Phys.: Condens.\ Matter {\bf 20}, 255251 (2008).

\bibitem{Bi:2008_PRB} 
R.~F.~Bishop, P.~H.~Y.~Li, R.~Darradi, J.~Schulenburg, and J.~Richter,
Phys.\ Rev.\ B {\bf 78}, 054412 (2008).

\bibitem{Orbach:1958_BetheAns}
R.~Orbach, Phys. Rev. {\bf 112}, 309 (1958);
C.~N.~Yang and C.~P.~Yang, {\it ibid.} {\bf 150}, 321 (1966); {\bf 150}, 327 (1966);
R~J.~Baxter, J.\ Stat.\ Phys. {\bf 9}, 145 (1973).

\bibitem{Ze:1998} 
C.~Zeng, D.~J.~J.~Farnell, and R.~F.~Bishop, 
J.\ Stat.\ Phys.\ {\bf 90}, 327 (1998).

\bibitem{Kr:2000} 
S.~E.~Kr{\"{u}}ger, J.~Richter, J.~Schulenburg, D.~J.~J.~Farnell, and R.~F.~Bishop, 
Phys.\ Rev.\ B {\bf 61}, 14607 (2000).

\bibitem{Fa:2001}
D.~J.~J.~Farnell, R.~F.~Bishop, and K.~A.~Gernoth,
Phys.\ Rev.\ B {\bf 63}, 220402(R) (2001).

\bibitem{Da:2005} 
R.~Darradi, J.~Richter, and D.~J.~J.~Farnell, 
Phys.\ Rev.\ B {\bf 72}, 104425 (2005).

\bibitem{Schm:2006} 
D.~Schmalfu$\ss$, R.~Darradi, J.~Richter, J.~Schulenburg, and D.~Ihle, 
Phys.\ Rev.\ Lett.\ {\bf 97}, 157201 (2006).
  
\bibitem{Bishop:2010_KagomeSq}
R.~F.~Bishop, P.~H.~Y.~Li, D.~J.~J.~Farnell, and C.~E.~Campbell,
Phys.\ Rev.\ B {\bf 82}, 104406 (2010).

\bibitem{ccm} 
We use the program package ``Crystallographic Coupled Cluster Method'' (CCCM) 
of D.~J.~J.~Farnell and J.~Schulenburg, see 
http://www-e.uni-magdeburg.de/jschulen/ccm/index.html.

\bibitem{Fa:2008} 
D.~J.~J.~Farnell and R.~F.~Bishop,
Int.\ J.\ Mod.\ Phys.\ B {\bf 22}, 3369 (2008).

\bibitem{Mo:1953}
P.~M.~Morse and H.~Feshbach,
{\it Methods of Theoretical Physics}, Part II (McGraw-Hill, New York, 1953).

\bibitem{Ha:1992}
C.~J.~Hamer, Zheng Weihong, and P.~Arndt,
Phys.\ Rev.\ B {\bf 46}, 6276 (1992).

\bibitem{Zh:1991}
Zheng Weihong, J.~Oitmaa, and C.~J.~Hamer,
Phys.\ Rev.\ B {\bf 43}, 8321 (1991).

\bibitem{Wh:1993}
S.~R.~White and D.~A.~Huse,
Phys.\ Rev.\ B {\bf 48}, 3844 (1993).

\bibitem{Nakatsuji:2005}
S.~Nakatsuji, Y.~Nambu, H.~Tonomura, O.~Sakai, S.~Jonas, C.~Broholm,
H.~Tsunetsugu, Y.~Qiu, and Y.~Maeno,
Science {\bf 309}, 1697 (2005).

\bibitem{Cheng:2011}
J.~G.~Cheng, G.~Li, L.~Balicas, J.~S.~Zhou, J.~B.~Goodenough, Cenke~Xu, 
and H.~D.~Zhou,
Phys.\ Rev.\ Lett.\ {\bf 107}, 197204 (2011).

\bibitem{Bhatta:2006}
S.~Bhattacharjee, V.~B.~Shenoy, and T.~Senthil, 
Phys.\ Rev.\ B {\bf 74}, 092406 (2006).

\bibitem{Lauchli:2006}
A.~Lauchli, F.~Mila, and K.~Penc, 
Phys.\ Rev.\ Lett.\ {\bf 97}, 087205 (2006).

\bibitem{Tsunetsugu:2006}
H.~Tsunetsugu and M.~Arikawa, 
J.\ Phys.\ Soc.\ Jpn.\ {\bf 75}, 083701 (2006).

\bibitem{Stoudenmire:2009}
E.~M.~Stoudenmire, S.~Trebst, and L.~Balents, 
Phys.\ Rev.\ B {\bf 79}, 214436 (2009).

\bibitem{Xu:2011}
C.~Xu, F.~Wang, Y.~Qi, L.~Balents, and M.~P.~A.~Fisher,
arXiv:1110.3328v1 [cond-mat.str-el] (2011).

\bibitem{FBLRC:2011} 
D.~J.~J.~Farnell, R.~F.~Bishop, P.~H.~Y.~Li, J.~Richter, and C.~E.~Campbell,
Phys.\ Rev.\ B {\bf 84}, 012403 (2011).

\bibitem{Xian:1994}
Y.~Xian,
J.\ Phys.: Condens.\ Matter {\bf 6}, 5965 (1994).

\bibitem{Farnell:2009}
D.~J.~J.~Farnell, J.~Richter, R.~Zinke and R.~F.~Bishop,
J.\ Stat.\ Phys.\ {\bf 135}, 175 (2009).

\bibitem{SS:1981}
B.~S.~Shastry and B.~Sutherland,
Physica B {\bf 108}, 1069 (1981).


\end{thebibliography}
%\begin{thebibliography}{}
%Author, \textit{Book title} (Publisher, place year) page numbers
% etc
%\end{thebibliography}

\clearpage

\end{document}